\documentclass[]{spie}  

\usepackage{amsmath,amsfonts,amssymb}
\usepackage{bm}
\usepackage{gensymb}
\usepackage{graphicx}
\usepackage{placeins}
\usepackage[colorlinks=true,allcolors=blue]{hyperref}
\usepackage{tabularx}
\usepackage{adjustbox}
\usepackage{array}
\usepackage{subcaption}
\usepackage{booktabs}
\usepackage{caption}
\usepackage{ragged2e}
\usepackage[numbers]{natbib}
\usepackage{xcolor}

\DeclareCaptionFormat{justified}{\justifying#1#2#3\par}
\title{Enabling Quantitative Polarimetry for Keck/NIRC2: Preliminary Mueller Matrix Model Calibration}

\author[1]{Manxuan Zhang$^*$}
\author[1]{Briley L. Lewis}
\author[1]{Maxwell A. Millar-Blanchaer}
\author[3]{Eduardo Marin}
\author[4]{Jayke S. Nguyen}
\author[5]{William Melby}
\author[3]{Carlos Alvarez}
\author[1]{Jaren N. Ashcraft}
\author[3]{Mahawa Cisse}
\author[6]{Charles-Antoine Claveau}
\author[3]{Jacques-Robert Delorme}
\author[3]{Greg Doppmann}
\author[2]{Michael P. Fitzgerald}
\author[6]{Matthew Freeman}
\author[3]{Percy Gomez}
\author[7,3]{Trisha Hammen}
\author[1]{Ryan Hersey}
\author[8]{Nemanja Jovanovic}
\author[3]{Marc Kassis}
\author[3]{Scott Lilley}
\author[6,9]{Jessica Lu}
\author[3]{James E. Lyke}
\author[8]{Keith Matthews}
\author[8,10]{Dimitri Mawet}
\author[1]{Thomas McIntosh}
\author[3]{Max Service}
\author[3]{Lauren Simmons}
\author[3]{Jacob Taylor}
\author[11]{Rob G. van Holstein}
\author[3]{Ed Wetherell}

\affil[1]{Department of Physics, University of California Santa Barbara, Santa Barbara, CA 93106 USA}
\affil[2]{Department of Physics and Astronomy, University of California Los Angeles, Los Angeles, CA 90024 USA}
\affil[3]{W.M. Keck Observatory, Kamuela, HI 96743 USA}
\affil[4]{Department of Astronomy, University of California San Diego, La Jolla, CA 92093 USA}
\affil[5]{Wyant College of Optical Sciences, University of Arizona, Tucson, AZ 85721 USA}
\affil[6]{Department of Astronomy, University of California Berkeley, Berkeley, CA 94720 USA}
\affil[7]{Department of Electrical and Computer Engineering, Colorado State University, Fort Collins, CO 80523 USA}
\affil[8]{California Institute of Technology, Pasadena, CA 91125 USA}
\affil[9]{Space Sciences Laboratory, University of California, Berkeley, CA 94720 USA}
\affil[10]{NASA Jet Propulsion Lab, Pasadena, CA 91109 USA}
\affil[11]{European Southern Observatory, Vitacura, Santiago, Chile}

\authorinfo{* Further author information: (Send correspondence to Manxuan Zhang)\\
E-mail: manxuanzhang@ucsb.edu}

\begin{document}

\maketitle

\begin{abstract}
The Keck/NIRC2 infrared imager was upgraded in 2025 with dual-beam polarimetric observing modes spanning approximately 1.1--4.1~$\mu$m ($JHKL'$ bands). We present a preliminary $JHK$ calibration of NIRC2 Polarimetry using a wavelength-dependent Mueller matrix model of the Keck tertiary mirror (M3), half-wave plate (HWP), image rotator (IMR), downstream optics, and Wollaston prism. We constrain the model downstream of M3 using dome flat sequences spanning ten HWP and nine IMR angles in each band. The Markov chain Monte Carlo (MCMC) maximum a posteriori (MAP) IMR retardances are $0.516_{-0.008}^{+0.008}$, $0.476_{-0.007}^{+0.009}$, and $0.483_{-0.005}^{+0.005}$ waves in $J$, $H$, and $K$, respectively; the HWP retardances are $0.54_{-0.02}^{+0.04}$, $0.44_{-0.05}^{+0.02}$, and $0.45_{-0.02}^{+0.02}$ waves; the effective dome-flat M3 diattenuations are $0.0197_{-0.0005}^{+0.001}$, $0.0186_{-0.0004}^{+0.002}$, and $0.0122_{-0.0001}^{+0.0003}$. Although the model reproduces the dominant modulation, the residuals show structure dependent on HWP and IMR angle. Measurement matrix inversion of unpolarized standard star observations gives M3 diattenuations of $0.0119\pm0.0009$, $0.0098\pm0.0004$, and $0.0068\pm0.0005$ in $J$, $H$, and $K_p$, substantially closer to Fresnel predictions for aluminum than the values derived from dome flats. The larger dome flat modulation may indicate polarization in the incident dome illumination or Mueller matrix model inaccuracies. These results establish an initial calibration framework while motivating improved input polarization constraints, fixed HWP parameters from previous laboratory measurements, model validation with polarized standard stars, and extension to $L'$.


\end{abstract}

\section{Introduction}
\label{sec:introduction}

Polarimetry provides access to information that is inaccessible from total intensity imaging alone. For example, in the optical and near-infrared, scattered light from dust grains is linearly polarized, while direct stellar light is often weakly polarized or unpolarized. This contrast between polarized circumstellar material and the stellar point-spread function (PSF) enables polarimetric differential imaging (PDI), a powerful observing strategy for high-contrast studies of protoplanetary disks, debris disks, the envelopes of evolved stars, and other extended scattering structures \citep{tinbergen1996astronomical,follette2023introduction}. 
Outside of high-contrast imaging, infrared polarimetry is also relevant for studies of solar system bodies, the galactic center, and magnetic field geometries on stars.


Despite its scientific utility, diffraction-limited infrared polarimetry remains comparatively rare on large 8--10m telescopes, particularly at thermal infrared wavelengths and in combination with spectroscopy, high-contrast coronagraphy, and laser guide star capabilities. Existing infrared polarimeters fed by adaptive optics (AO) systems include VLT/SPHERE/IRDIS \citep{de2020polarimetric,van2020polarimetric}, Gemini/GPI(2.0) \citep{perrin2010imaging,millar2016gpi,chilcote2020gpi}, Subaru/SCExAO/CHARIS \citep{gj2021full}, and Subaru/IRCS \citep{watanabe2018near, terada2018thermal}. These instruments have demonstrated the value of near-infrared polarimetry for a variety of science cases, but most of these systems (with the exception of Subaru/IRCS) operate only through $K$ band. Thermal-infrared polarimetry (i.e. $L'$ band) remains much less common, though longer wavelengths can probe larger grains, warmer material, and embedded structures that are complementary to shorter wavelength near-infrared images. A newly commissioned polarimetric mode on Keck/NIRC2 therefore fills an important observational niche by combining polarimetry with the angular resolution of a 10m telescope, broad near- to thermal-infrared wavelength coverage, and extensive high-contrast observing infrastructure already available for Keck II AO.

NIRC2 is the facility near-infrared camera on the Keck II telescope and operates behind the Keck adaptive optics system over $\sim$ 1--5~$\mu$m \citep{van2004performance,wizinowich2006wm,lilley2024keck}. The instrument supports a wide range of imaging and spectroscopic modes, including broadband and narrowband filters, pupil masks, grisms, slits, Lyot coronagraphs, and vector vortex coronagraphs optimized for $K$ and $L'/M$ bands \citep{castella2016commissioning,xuan2018characterizing}. NIRC2 has been used extensively for high angular resolution studies of exoplanets and brown dwarfs, circumstellar disks, the Galactic center, solar system objects, and precision astrometry. A recent upgrade to NIRC2 expanded the instrument's capabilities to include imaging polarimetry, coronagraphic imaging polarimetry, and low-to-medium resolution spectropolarimetry.

The NIRC2 Polarimetry upgrade (referred to as NIRC2-Pol) created dual-channel IR polarimetry capabilities in the instrument with the installation of a Wollaston prism and 5 x 10'' field mask in NIRC2 and two selectable half-wave plate (HWP) modulators on the Keck II adaptive optics bench \cite{lewis2026nirc}.  The Wollaston prism separates the incoming beam into two orthogonal linear polarization states and the field mask prevents overlap of the ordinary and extraordinary beams on the detector, while the HWP modulates the incident polarization angle so that Stokes $I$, $Q$, and $U$ can be recovered from measurements at multiple HWP angles. Two different HWPs are used, one for the shorter-wavelength $J/H/K$ bands and one for $L'$ band, providing polarimetric coverage from approximately 1.1 to 4.1~$\mu$m. The Wollaston prism was installed in 2019, while the HWP hardware and associated optomechanics were completed later as part of the Keck II Precision Calibration Unit upgrade in 2025. Laboratory characterization of HWP retardance and throughput was performed prior to on-sky commissioning using a dual-rotating retarder polarimeter \citep{melby2024half}. NIRC2-Pol commissioning and science verification observations were subsequently obtained in 2025--2026, demonstrating the new mode on sky in multiple observing configurations.


Realizing the full scientific potential of NIRC2-Pol requires careful calibration of instrumental polarization and polarization cross-talk. This requirement is especially important for a Nasmyth-mounted telescope and instrument system such as Keck/NIRC2, where reflection off the tertiary/Nasmyth mirror (hereafter referred to as M3) induces altitude-dependent instrumental polarization. As the HWP is located downstream of M3, polarization effects introduced by the telescope are not removed by polarimetric differential imaging alone. Additionally, isolating the effect of the instrumental polarization induced by M3 is not possible without modeling the effect of downstream polarimetric components. Accurate quantitative polarimetry therefore requires a Mueller matrix model of the full optical train. This model includes, from upstream to downstream: M3, the HWP, the image rotator (IMR), intermediate optics, and the Wollaston prism. Similar Mueller matrix models have been built for other high-contrast polarimeters to correct instrumental polarimetric effects and recover astrophysical Stokes parameters with improved accuracy \citep{de2020polarimetric,van2020polarimetric,gj2021full,zhang2024matrix_inversion, mcintosh2026charis}.

In this work, we present progress on the calibration and instrument characterization for NIRC2-Pol. We describe the instrument configuration and construction of a multi-wavelength Mueller matrix model in \textit{JHK} bands in Section \ref{sec:instrument_model}. We then use both dome flat data and unpolarized standard stars observed over a range of telescope altitudes to constrain the instrumental polarization produced by the Keck tertiary mirror in Section \ref{sec:calibration_data}. We also discuss practical calibration strategies that can improve upon current progress, including new HWP and IMR angles during dome flat calibrations, more polarimetric parameters modelled, and verification of the current model with suitable polarized standards.


\section{NIRC2-Pol Instrument and Polarimetric Model}
\label{sec:instrument_model}

\subsection{Instrument Configuration and Observing Modes}
\label{subsec:instrument_configuration}

\begin{figure}[hbtp!]
    \centering
    \IfFileExists{NIRC2_Diagram_No_M3.png}{%
        \includegraphics[width=\textwidth]{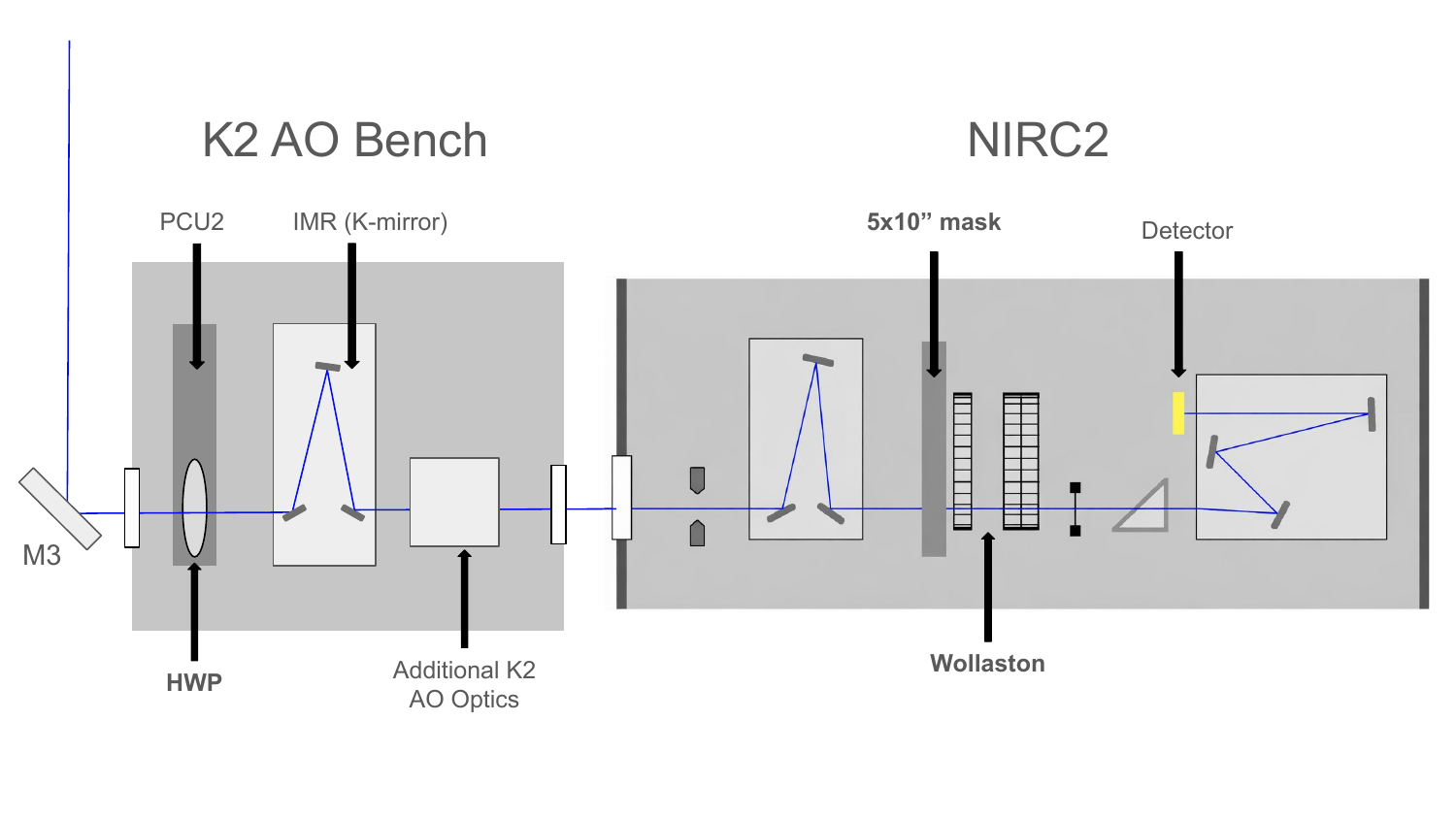}%
    }{%
        \fbox{\parbox[c][2.0in][c]{0.95\textwidth}{\centering
        Figure file needed: \texttt{NIRC2\_Diagram\_No\_M3.png}}}%
    }
    \caption{Simplified diagram of the Keck II adaptive-optics bench and NIRC2 highlighting optics added for the upgrade in bold. The addition of an HWP on the adaptive-optics bench and a Wollaston prism and 5x10" field mask in NIRC2 enables dual-beam polarimetry. M3, the image rotator (IMR/K-mirror), and other intermediate optics are also relevant for the Mueller matrix modeling in this work.}
    \label{fig:nirc2_pol_diagram}
\end{figure}

A simplified diagram of the beam path from the HWP to the NIRC2 detector is shown in Fig.~\ref{fig:nirc2_pol_diagram}. NIRC2-Pol is designed as a dual-channel polarimeter, where the Wollaston prism splits light into an ordinary and extraordinary beam. A 5" x 10" field mask is placed before the Wollaston during polarimetric observations to avoid overlap between the ordinary and extraordinary beams. Rotation of the HWP exchanges the sign of the linear Stokes signal measured in the two beams, enabling retrieval of Stokes \textit{I}, \textit{Q}, and \textit{U} and differential suppression of common-mode atmospheric and instrumental effects. Two separate HWPs are used for JHK and L' respectively, with a manual daytime swap necessary to switch between them. More information about the instrument and its possible configurations can be found in Lewis et al. 2026 \citep{lewis2026nirc}.

\subsection{Polarization Conventions and Definitions}
\label{subsec:polarization_conventions}

The polarization state is represented by the Stokes vector
\begin{equation}
    \boldsymbol{S} =
    \begin{pmatrix}
        I \\
        Q \\
        U \\
        V
    \end{pmatrix},
    \label{eq:stokes_vector}
\end{equation}
where $I$ is total intensity, $Q$ and $U$ describe orthogonal linear-polarization states, and $V$ describes circular polarization. The normalized Stokes parameters are
\begin{equation}
    q = \frac{Q}{I}, \qquad
    u = \frac{U}{I}, \qquad
    v = \frac{V}{I}.
    \label{eq:normalized_stokes}
\end{equation}
The degree of linear polarization (DoLP) and angle of linear polarization (AoLP) are
\begin{equation}
    DoLP = \sqrt{q^2+u^2},
    \label{eq:dolp_definition}
\end{equation}
and
\begin{equation}
    AOLP = \frac{1}{2}\arctan2(\frac{u}{q}),
    \label{eq:aolp_definition}
\end{equation}
respectively, before applying the instrument-to-sky coordinate rotation.

We distinguish between the sky Stokes frame and the NIRC2 instrument frame. In the
sky frame, displayed with north up and east to the left, $+Q_{\rm sky}$ denotes
north--south linear polarization and $-Q_{\rm sky}$ denotes east--west linear
polarization. Likewise, $+U_{\rm sky}$ denotes polarization at $+45^\circ$ from
north toward east (the northeast--southwest axis), while $-U_{\rm sky}$ denotes
the orthogonal northwest--southeast axis. The final on-sky AoLP is measured east of north. The parallactic angle is defined from celestial north to zenith. 

In the NIRC2 instrument frame, the horizontal detector direction is identified
with $+Q_{\rm inst}$ and the vertical detector direction with $-Q_{\rm inst}$.
Angles assigned to the HWP and image rotator are defined while looking from the
telescope side toward NIRC2 (i.e. along the direction of beam propagation). A
positive mechanical angle is counterclockwise in this view. This convention is
the same for the HWP and the image rotator. 

For the current data reduction, single differences are formed by subtracting flux in the bottom Wollaston beam from the top Wollaston beam. Future work is necessary to distinguish between the ordinary and extraordinary beams of the Wollaston. The corresponding normalized beam
difference is
\begin{equation}
 ND =\frac{f_{\rm top}-f_{\rm bottom}}
 {f_{\rm top}+f_{\rm bottom}}.
\label{eq:normalized_difference}
\end{equation}
An initial coordinate validation test with on-sky L' data of AB Aurigae used these coordinate definitions and an empirical HWP offset of $\theta_{\rm offset}=-8^\circ$, obtained by fitting a $0 - 180^{\circ}$
HWP scan with a term proportional to
$\sin[4(\theta_{\rm HWP}-\theta_{\rm offset})]$ \citep{lewis2026nirc}.  Changing the Wollaston beam subtraction order to top flux subtracted from bottom flux is equivalent to shifting the adopted HWP offset by $45^\circ$.

\subsection{Mueller-Matrix Descriptions}
\label{subsec:mueller_description}

A Mueller matrix ($M$) transforms an incident Stokes vector ($\boldsymbol{S}_{\rm in}$) according to
\begin{equation}
    \boldsymbol{S}_{\rm out} = M\boldsymbol{S}_{\rm in}.
    \label{eq:mueller_transform}
\end{equation}
For a component with diattenuation $\epsilon$ and retardance $\phi$ (assuming perfect transmission and no depolarization), we use:
\begin{equation}
M_{\rm comp} =
\begin{pmatrix}
1 & \epsilon & 0 & 0 \\
\epsilon & 1 & 0 & 0 \\
0 & 0 & \cos\phi\sqrt{1-\epsilon^2} & \sin\phi\sqrt{1-\epsilon^2} \\
0 & 0 & -\sin\phi\sqrt{1-\epsilon^2} & \cos\phi\sqrt{1-\epsilon^2}
\end{pmatrix}.
\label{eq:diattenuating_retarder}
\end{equation}
To rotate optical components, we use the rotation matrix:
\begin{equation}
T(\theta) =
\begin{pmatrix}
1 & 0 & 0 & 0 \\
0 & \cos 2\theta & \sin 2\theta & 0 \\
0 & -\sin 2\theta & \cos 2\theta & 0 \\
0 & 0 & 0 & 1
\end{pmatrix}.
\label{eq:rotation_matrix}
\end{equation}
A rotated optical component is represented by $T(-\theta)M_{\rm comp}T(\theta)$.

For a given HWP angle, IMR angle, telescope altitude, parallactic angle, and wavelength, the detector Stokes vector can be represented schematically as
\begin{equation}
\begin{aligned}
\boldsymbol{S}_{\rm det} ={}&
M_{\rm Woll}
T(-\theta_{\rm opt})M_{\rm opt}T(\theta_{\rm opt})
T(\theta_{\rm IMR})M_{\rm IMR}T(-\theta_{\rm IMR}) \\
&\times T(\theta_{\rm HWP})M_{\rm HWP}T(-\theta_{\rm HWP}) T(a)M_{\rm M3}T(p) \boldsymbol{S}_{\rm in}
\end{aligned}
\label{eq:full_mueller_chain}
\end{equation}
where $M_{\rm M3}$ represents the Keck tertiary mirror, $M_{\rm HWP}$ represents the selected HWP, $M_{\rm IMR}$ represents the NIRC2 image rotator, $M_{\rm opt}$ represents additional static downstream optics, and $M_{\rm Woll}$ represents the Wollaston analyzer. The angles $a$ and $p$ denote telescope altitude and parallactic angle respectively. Negative rotations are applied to the HWP and IMR angles as their rotation is in the negative direction. The model is evaluated independently in each band because the retardance and diattenuation of both transmissive and reflective elements vary with wavelength.

Optical elements that remain fixed relative to one another and share a common reference frame can be represented by a single effective Mueller matrix. The HWP and IMR are modeled separately as they rotate independently, and M3 is represented with its own Mueller matrix as its polarization axes are defined by its plane of incidence in the telescope reference frame rather than by the NIRC2 instrument frame. The fitted parameters presently include M3 diattenuation, HWP retardance, IMR retardance, and an effective combined retardance and orientation for the static downstream optics between the IMR and Wollaston.  Diattenuation from the HWP through the static optics is currently assumed to be strongly suppressed by beam and HWP angle differencing, and the Wollaston prism is treated as an ideal polarizing beamsplitter. Previous laboratory dual-rotating-retarder polarimeter measurements provide additional verification of fitted parameters on the expected retardance of the $J/H/K$ and $L'$ HWPs \citep{melby2024half}, while the theoretical diattenuation from aluminum is used as a point of comparison for fitted M3 diattenuation.

\subsection{Measured Beam Differences and Model Inversion}
\label{subsec:measurement_inversion}

For each exposure $i$, obtained at a unique combination of HWP angle, IMR
angle, wavelength, and, for on-sky observations, telescope altitude and
parallactic angle, we measure fluxes $f_{{\rm top},i}$ and
$f_{{\rm bottom},i}$ in the two Wollaston outputs.  The normalized single difference for exposure $i$, denoted $ND_i$, is calculated from $f_{{\rm top},i}$ and $f_{{\rm bottom},i}$ using the top-minus-bottom convention defined in Eq.~\ref{eq:normalized_difference}. 
The uncertainty on $ND_i$ is obtained by propagating the photometric uncertainties of the two beam measurements.  The error model used for the dome-flat MCMC fits is specified explicitly in Section~\ref{subsec:photometric_errors_centering}.

For dome-flat, twilight-sky, and empty-sky data, square apertures that contain
the majority of each Wollaston output are used.  For standard-star
observations, circular apertures with fixed radii for each target contain
approximately $95\%$ of the point-spread-function (PSF) flux in each
Wollaston beam.

On-sky sequences use the four standard HWP angles
\begin{equation}
    \theta_{\rm HWP}
    =0^\circ,\ 45^\circ,\ 22.5^\circ,\ 67.5^\circ,
    \label{eq:standard_hwp_angles}
\end{equation}
where the first pair measures the two signs of Stokes $Q$ and the second pair
measures the two signs of Stokes $U$ in the instrument frame. For each HWP
angle, the normalized single difference $ND^{\text{HWP} \theta}$ is calculated using the top-minus-bottom convention defined in
Eq.~\ref{eq:normalized_difference}. The normalized instrument-frame Stokes
parameters are then
\begin{equation}
    q_{\rm inst}
    =
    \frac{1}{2}
    \left(
    ND^{0^\circ}-ND^{45^\circ}
    \right),
    \label{eq:q_normalized_double_difference}
\end{equation}
and
\begin{equation}
    u_{\rm inst}
    =
    \frac{1}{2}
    \left(
    ND^{22.5^\circ}-ND^{67.5^\circ}
    \right).
    \label{eq:u_normalized_double_difference}
\end{equation}

For dome flat and twilight sky calibration measurements, both the HWP and IMR are modulated to determine how their polarimetric properties change with angle; this is crucial because neither component is static during typical on-sky observations. The physical polarimetric parameters are therefore fitted directly to the normalized single differences, rather than first collapsing the calibration sequence into $Q_{\rm inst}$ and $U_{\rm inst}$.  For unpolarized standard stars, the Stokes vector entering the HWP is recovered using the calibrated downstream measurement matrix.  For polarized standards and science targets, the incident on-sky Stokes vector is recovered using the full Mueller chain in Eq.~\ref{eq:full_mueller_chain} and a measurement matrix inversion procedure described in Section \ref{subsec:m3_fit_method}.  

\section{Internal Calibration}
\label{sec:calibration_data}

Quantitative recovery of astrophysical Stokes parameters requires a Mueller
matrix that includes both telescope and instrument terms.  We constructed the
current $JHK$ model from dome flat calibration sequences and will ultimately
use the downstream instrument Mueller matrix to determine M3 mirror parameters and total polarimetric accuracy with on-sky observations of unpolarized and polarized
standard stars respectively.  

\subsection{Calibration Sequence Design}
\label{subsec:calibration_sequence}

No dedicated polarized calibration source is available upstream of the HWP on
the Keck II adaptive optics bench.  We therefore used dome flats as a bright,
repeatable source for the $JHK$ calibration.  The incident dome illumination
was treated as unpolarized before M3, and the reflection from M3 was assumed to be the only component generating a small linear polarization signal.  The DoLP observed from the amplitude is only approximately $0.5$--$2\%$ in $JHK$. With such a low incident DoLP, the polarimetric modulation signal that constrains the downstream retarders is limited.

For each band, the instrument calibration sequence sampled HWP angles from
$0^\circ$ through $90^\circ$ in $10^\circ$ increments at each of nine image
rotator (IMR) angles from $0^\circ$ through $120^\circ$ in $15^\circ$ increments.  The resulting $10\times9=90$ measurements per band were used in the fits. The $K$-band integrations were longer, as the dome lamp was substantially fainter in $K$.  All data were obtained in CDS readout mode with two reads per exposure. Specific camera configurations for each wavelength band are detailed in Table \ref{tab:calibration_sequences}.

\begin{table}[t]
\centering
\caption{Dome-flat calibration sequences used for the $JHK$ NIRC2-Pol
Mueller-matrix fits.}
\label{tab:calibration_sequences}
\begin{tabular}{llccc}
\toprule
Band & Source & Date (UT) & $t_{\rm int}$ (s) & Coadds \\
\midrule
$J$ & Dome flat & 2026 Jan 27 & 10 & 1 \\
$H$ & Dome flat & 2026 Jan 27 & 10 & 1 \\
$K$ & Dome flat & 2026 Jan 27 & 45 & 1 \\
\bottomrule
\end{tabular}
\end{table}

\subsection{Photometric Uncertainties and Single Difference Offsets}
\label{subsec:photometric_errors_centering}

The normalized single difference for each i-th exposure $ND_i$, corresponds to a unique configuration of the instrument (unique HWP and IMR angle combination). $ND_i$ is defined in Eq.~\ref{eq:normalized_difference}. Initial fits to determine starting points for MCMC analysis using \texttt{scipy.minimize} did not include photometric uncertainties. For the MCMC analysis, the
top and bottom beam aperture sums were treated as independent Poisson counts, such that the variance on each aperture sum $\sigma^2$ is given by
\begin{equation}
    \sigma_{{\rm top},i}^2=\lvert f_{{\rm top},i}\rvert,
    \qquad
    \sigma_{{\rm bottom},i}^2=\lvert f_{{\rm bottom},i}\rvert .
\end{equation}
Analytical propagation through the normalized-difference definition gives
\begin{align}
 \sigma_{ND,i}^2 &=
 \left[
 \frac{2f_{{\rm bottom},i}}
 {(f_{{\rm top},i}+f_{{\rm bottom},i})^2}
 \right]^2\sigma_{{\rm top},i}^2 \nonumber\\
 &\quad+
 \left[
 \frac{-2f_{{\rm top},i}}
 {(f_{{\rm top},i}+f_{{\rm bottom},i})^2}
 \right]^2\sigma_{{\rm bottom},i}^2 .
 \label{eq:single_difference_uncertainty}
\end{align}
This error model does not include read noise or covariance between the two beam apertures.

At each fixed IMR angle, the normalized single differences measured across the HWP rotation sequence had a nonzero mean. Before fitting, this additive offset in the normalized single differences was removed by subtracting the mean normalized difference for every data point at a specific IMR angle. 
The \texttt{scipy.optimize.minimize} fits used the unweighted group mean, whereas the MCMC analysis used the inverse-variance weighted mean $\widehat{ND}_{g}$, where:


\begin{equation}
    \widehat{ND}_{g}=
    \frac{\sum_{i\in g}ND_i/\sigma_{ND,i}^{2}}
         {\sum_{i\in g}1/\sigma_{ND,i}^{2}} .
    \label{eq:weighted_imr_centering}
\end{equation}
The same method of removing the normalized single differences means was applied to the model when fitting for Mueller matrix parameters. This step of removing single difference offsets attempts to remove uneven flat-fielding effects and static diattenuation effects between the HWP and the Wollaston prism, as those effects are heavily suppressed with on-sky double differencing. However, this approach has important limitations. By calculating nine independent mean single difference offset levels, a real IMR dependent diattenuation signal can be erased. It can also weaken the ability to place constraints on any effects that change the mean offset at an IMR angle (e.g. IMR diattenuation, any static diattenuations in the beam path, Wollaston throughput ratio).  As the same mean is subtracted from every measurement within an IMR group, the centered measurements also now include correlated errors. The present likelihood treats the centered points as independent and does not include this covariance induced by IMR angle offset centering. The reported uncertainties are therefore conditional on this approximation.

\subsection{Bayesian MCMC Analysis}
\label{subsec:mcmc_analysis}

\subsubsection{Starting Guesses from \texttt{scipy.optimize.minimize}}
\label{subsec:scipy_fit}

To obtain a starting point for MCMC analysis, we first obtained a solution in each band using
the bounded Nelder-Mead implementation of
\texttt{scipy.optimize.minimize} \citep{scipy}. The input Stokes vector before M3 was fixed to $(I,Q,U,V)=(1,0,0,0)$. A $45^\circ$ altitude (telescope altitude during dome flats) rotation was applied immediately after M3 as per Eq. \ref{eq:full_mueller_chain}. Retardances are reported here in waves, with one wave corresponding to $2\pi$ radians of phase delay.

Five quantities were fitted: the retardance and rotation angle of an optics term encompassing all components between the IMR and the Wollaston prism, the IMR retardance, the HWP retardance, and the M3 diattenuation. The Wollaston transmission ratio was fixed to unity and the optics, IMR, and HWP diattenuations were fixed to zero. HWP fast-axis offsets were fixed to values calculated in the procedure described in Section \ref{subsec:polarization_conventions} with exact values and details on the data collected to determine these offsets shown in Table \ref{tab:hwp_fast_axis_offsets}. 

\begin{table*}[hbtp!]
\centering
\caption{Results of dome flat HWP modulation sequences used to determine the HWP fast-axis offsets used in the $JHK$ Mueller matrix fits. The telescope was parked at the dome flat position at an altitude of $45^\circ$ and an azimuth of $64^\circ$. The $J$- and $H$-band sequences sampled $0^\circ\leq\theta_{\rm HWP}\leq180^\circ$ in $10^\circ$ increments. The available $K$-band sequence sampled $0^\circ\leq\theta_{\rm HWP}\leq80^\circ$ at the same spacing. In each band, the normalized differences were fitted with $ND(\theta_{\rm HWP})=C+A\sin[4(\theta_{\rm HWP}-\theta_{\rm offset})]$, using the same definition of $\theta_{\rm offset}$ introduced in Section~\ref{subsec:polarization_conventions}. The reported uncertainties are the $1\sigma$ standard errors obtained from the diagonal element corresponding to $\theta_{\rm offset}$ in the \texttt{scipy.optimize.curve\_fit} parameter
covariance matrix.}
\label{tab:hwp_fast_axis_offsets}
\begin{tabular}{lccccc}
\toprule
Band & Date (UT) & HWP range & $t_{int}$ [s] & Coadds
& $\theta_{\rm off}$ [$^\circ$] \\
\midrule
$J$ & 2025 November 21 & $0^\circ$--$180^\circ$ & 40  & 1
& $17.9\pm0.1$ \\
$H$ & 2025 October 2   & $0^\circ$--$180^\circ$ & 30  & 1
& $19.8\pm0.3$ \\
$K$ & 2025 September 15 & $0^\circ$--$80^\circ$ & 150 & 1
& $22.8\pm0.2$ \\
\bottomrule
\end{tabular}
\end{table*}

The starting points for each band's minimization procedure are listed in Table~\ref{tab:scipy_starts}.  The $H$ and $K$ optimizations were started from parameters obtained in the preceding shorter wavelength's minimization best fit, while their IMR retardances were set to $0.25$ waves as the IMR was expected to deviate significantly from 0.5 waves at longer wavelengths. From Fresnel coefficients, the expected M3 diattenuations were $0.012033$, $0.008698$, and $0.007708$ in $J$, $H$, and $K$, respectively.  The $J$ fit began at its nominal M3 value, while the $H$ and $K$ fits instead started with the preceding band's fitted M3 value. 

\begin{table*}[hbtp!]
\centering
\caption{Starting points used for the \texttt{scipy.optimize.minimize} fits.  Only free
parameters are shown.}
\label{tab:scipy_starts}
\begin{tabular}{lccc}
\toprule
Parameter & $J$ & $H$ & $K$ \\
\midrule
Optics retardance [waves] & 0.00000 & 0.09419 & 0.06532 \\
Optics angle [$^\circ$] & 0.000 & $-65.803$ & $-64.113$ \\
IMR retardance [waves] & 0.50000 & 0.25000 & 0.25000 \\
HWP retardance [waves] & 0.50000 & 0.45768 & 0.46077 \\
M3 diattenuation & 0.012033 & 0.019712 & 0.018523 \\
\bottomrule
\end{tabular}
\end{table*}

The bounds supplied to the optimizer are listed in
Table~\ref{tab:scipy_bounds}. As no errors were passed to the minimization routine, the cost function minimized:
\begin{equation}
    \mathcal{C}(\boldsymbol{\theta})
      = \frac{1}{2}\sum_{i=1}^{90}
        \left[ND'_i-m'_i(\boldsymbol{\theta})\right]^2 ,
    \label{eq:scipy_objective}
\end{equation}
where $m'_i$ is the i-th model point, corresponding to the predicted normalized single difference for the i-th unique HWP and IMR angle combination. Each minimization pass was restarted from the preceding pass's solution. Iteration stopped when the
relative improvement in the single difference's root mean squared error was below $1\%$, with at most
50 passes allowed. This required four, two, and two passes for $J$, $H$, and $K$, respectively.  

\begin{table}[hbtp!]
\centering
\caption{Bounds on the free parameters in the \texttt{scipy.optimize.minimize} fits.}
\label{tab:scipy_bounds}
\begin{tabular}{lc}
\toprule
Parameter & Bound \\
\midrule
Optics retardance [waves] & $[0,1]$ \\
Optics angle [$^\circ$] & $[-90,90]$ \\
IMR retardance [waves] & $[0,1]$ \\
HWP retardance [waves] & $[0,1]$ \\
M3 diattenuation & $[0,0.05]$ \\
\bottomrule
\end{tabular}
\end{table}

\begin{table*}[hbtp!]
\centering
\caption{Final deterministic solutions from \texttt{scipy.optimize.minimize}.  These values supplied guidance for
the subsequent MCMC analysis but are not used as uncertainty estimates.}
\label{tab:scipy_results}
\begin{tabular}{lccc}
\toprule
Parameter & $J$ & $H$ & $K$ \\
\midrule
Optics retardance [waves] & 0.094195 & 0.065325 & 0.181722 \\
Optics angle [$^\circ$] & $-65.803$ & $-64.113$ & $-5.945$ \\
IMR retardance [waves] & 0.484108 & 0.480156 & 0.482910 \\
HWP retardance [waves] & 0.457681 & 0.460768 & 0.446580 \\
M3 diattenuation & 0.019712 & 0.018523 & 0.012222 \\
\bottomrule
\end{tabular}
\end{table*}


\subsubsection{Likelihood, Priors, and Initialization}
\label{subsubsec:mcmc_method}

We sampled the same five optical parameters together with an additive log scatter parameter, $\ln f$, using \texttt{emcee} \citep{emcee}.  Given the residual
$r_i=ND'_i-m'_i(\boldsymbol{\theta})$, the variance $s_i$ and log likelihood $\ln\mathcal{L}$ are:
\begin{align}
    s_i^2 &= \sigma_{ND,i}^2+\exp(2\ln f), \\
    \ln\mathcal{L}
      &= -\frac{1}{2}\sum_{i=1}^{90}
      \left(\frac{r_i^2}{s_i^2}+\ln s_i^2\right).
    \label{eq:mcmc_log_likelihood}
\end{align}

The addition of the additive log scatter parameter $\ln f$ has been used in earlier VAMPIRES Mueller matrix calibration and in models with incompletely known measurement uncertainties \citep{zhang2023characterizing,hogg2010fitting}. The propagated photometric uncertainties account for the estimated measurement noise but do not describe additional scatter arising from unmodeled polarimetric effects, calibration systematics, or other variations in the data. Using the photometric uncertainties alone could therefore produce overly restrictive posterior intervals.

\begin{table}[hbtp!]
\centering
\caption{Shared MCMC priors.  All priors were uniform within the listed interval and zero outside it. $\ln f$ is an additive log scatter parameter.}
\label{tab:mcmc_priors}
\begin{tabular}{lc}
\toprule
Parameter & Prior Interval \\
\midrule
Optics retardance [waves] & $\mathcal{U}(0,1)$ \\
Optics angle [$^\circ$] & $\mathcal{U}(-45,45)$ \\
IMR retardance [waves] & $\mathcal{U}(0,1)$ \\
HWP retardance [waves] & $\mathcal{U}(0,1)$ \\
M3 diattenuation & $\mathcal{U}(0,0.05)$ \\
$\ln f$ & $\mathcal{U}(-10,0)$ \\
\bottomrule
\end{tabular}
\end{table}

Table~\ref{tab:mcmc_priors} gives the shared, independent uniform priors. The \texttt{scipy.optimize.minimize} best fit results in Table \ref{tab:scipy_results} were used as the starting points for the MCMC runs. The optics angle bounds were restricted to $[-45^\circ,45^\circ]$ in order to avoid any angular degeneracies. Any starting value outside that interval was wrapped into an equivalent value in the $[-45^\circ,45^\circ]$ range.  Each run used fifteen walkers and the default \texttt{emcee} stretch move with scale parameter $a=2$.  Walkers were initialized with small Gaussian perturbations about the tabulated center.  In
the internal parameterization, the perturbation scale in each dimension was the smaller of $10^{-3}$ and $10^{-4}$ times the prior width, with a floor of
$10^{-8}$. 


The first ensembles located substantially higher-density regions far from some of their original centers, especially in $\ln f$ and, for the $J$ and $H$, optics rotation angle.  We therefore started new, independent MCMC runs centered on the highest likelihood log-posterior chains from each initial run.  The second MCMC run start values are listed in Table~\ref{tab:mcmc_restart_starts}. 

\begin{table*}[hbtp!]
\centering
\caption{Highest likelihood log-posterior MCMC restart values.}
\label{tab:mcmc_restart_starts}
\begin{tabular}{lccc}
\toprule
Parameter & $J$ & $H$ & $K$ \\
\midrule
Optics retardance [waves] & 0.095133 & 0.039674 & 0.178862 \\
Optics angle [$^\circ$] & 24.014 & $-44.977$ & $-6.048$ \\
IMR retardance [waves] & 0.515066 & 0.478353 & 0.483647 \\
HWP retardance [waves] & 0.542945 & 0.440510 & 0.447779 \\
M3 diattenuation & 0.019614 & 0.018589 & 0.012213 \\
\bottomrule
\end{tabular}
\end{table*}


\subsubsection{Posterior Results and Convergence}
\label{subsubsec:mcmc_results}




\begin{table*}[hbtp!]
\centering
\caption{Final posterior medians and MAP estimates. Errors are the 16th-84th percentile posterior distribution values about the median.}
\label{tab:mcmc_median_map}
\begin{tabular}{llccc}
\toprule
Parameter & Estimator & $J$ & $H$ & $K$ \\
\midrule
Optics retardance [waves]
 & Median & $0.10_{-0.01}^{+0.03}$ & $0.03_{-0.01}^{+0.01}$ & $0.18_{-0.04}^{+0.05}$ \\
 & MAP & $0.10_{-0.01}^{+0.03}$ & $0.04_{-0.01}^{+0.01}$ & $0.18_{-0.04}^{+0.05}$ \\
\addlinespace
Optics angle [$^\circ$]
 & Median & $20_{-9}^{+10}$ & $-42_{-3}^{+9}$ & $-6_{-4}^{+2}$ \\
 & MAP & $20_{-9}^{+10}$ & $-45_{-3}^{+9}$ & $-6_{-4}^{+2}$ \\
\addlinespace
IMR retardance [waves]
 & Median & $0.514_{-0.008}^{+0.008}$ & $0.482_{-0.007}^{+0.009}$ & $0.485_{-0.005}^{+0.005}$ \\
 & MAP & $0.516_{-0.008}^{+0.008}$ & $0.476_{-0.007}^{+0.009}$ & $0.483_{-0.005}^{+0.005}$ \\
\addlinespace
HWP retardance [waves]
 & Median & $0.56_{-0.02}^{+0.04}$ & $0.42_{-0.05}^{+0.02}$ & $0.45_{-0.02}^{+0.02}$ \\
 & MAP & $0.54_{-0.02}^{+0.04}$ & $0.44_{-0.05}^{+0.02}$ & $0.45_{-0.02}^{+0.02}$ \\
\addlinespace
M3 diattenuation
 & Median & $0.0196_{-0.0005}^{+0.001}$ & $0.0188_{-0.0004}^{+0.002}$ & $0.0123_{-0.0001}^{+0.0003}$ \\
 & MAP & $0.0197_{-0.0005}^{+0.001}$ & $0.0186_{-0.0004}^{+0.002}$ & $0.0122_{-0.0001}^{+0.0003}$ \\
\addlinespace
$\ln f$
 & Median & $-6.21_{-0.08}^{+0.08}$ & $-6.66_{-0.07}^{+0.08}$ & $-7.56_{-0.07}^{+0.08}$ \\
 & MAP & $-6.25_{-0.08}^{+0.08}$ & $-6.70_{-0.07}^{+0.08}$ & $-7.60_{-0.07}^{+0.08}$ \\
\bottomrule
\end{tabular}
\end{table*}

\begin{figure}[hbtp!]
\includegraphics[width=\linewidth]{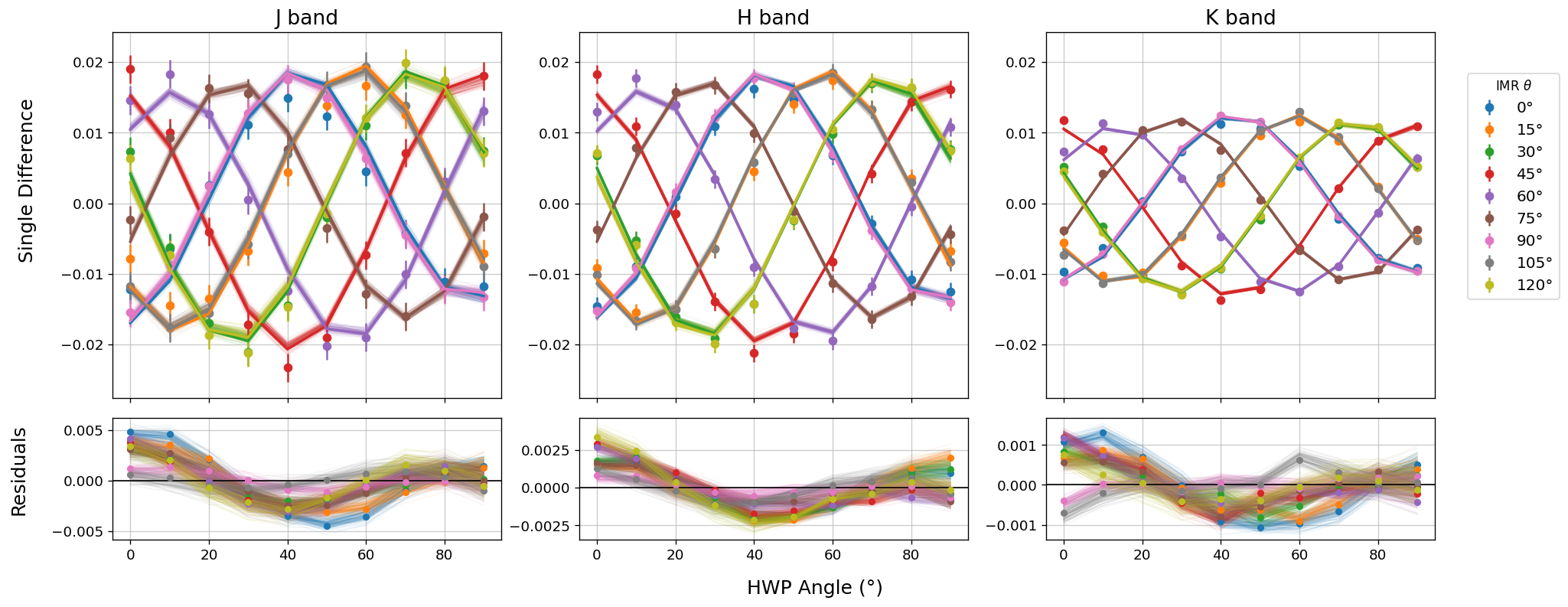}

\caption{One hundred random posterior draws fit to normalized single differences in the $J$, $H$, and $K$ bands. The upper panels show the measured values with uncertainties given by the additive scatter parameter $\ln f$. The lower panels show the corresponding residuals. The models reproduce the dominant HWP and IMR dependent modulation, but residual structure remains correlated with HWP angle. Single differences and residuals are largest in $J$ and smallest in $K$.}
\label{fig:JHK_random_chains}
\end{figure}

Table~\ref{tab:mcmc_median_map} reports both maximum-a-posteriori (MAP) and median parameter estimates, while the autocorrelation-time diagnostics are listed in Table~\ref{tab:mcmc_tau}. For each band, the integrated autocorrelation time was estimated separately for every fitted parameter using the full saved chain. Following the standard \texttt{emcee} heuristic, the burn-in was set to twice the largest autocorrelation time, thereby using the slowest-mixing parameter to determine how much of the beginning of the chain to discard. The thinning interval was set to half the smallest autocorrelation time to reduce the size of the posterior sample without thinning more aggressively than warranted by the fastest mixing parameter. This gave burn-in and thinning values of $(7688,96)$ steps for $J$, $(18065,118)$ for $H$, and $(5889,90)$ for $K$. After these cuts, the flattened posteriors contained 53,940, 62,760, and 72,885 retained walker samples, respectively. The lowest chain length to autocorrelation time ratios were 91.8 for $J$, 56.7 for $H$, and 150.5 for $K$, in each case for M3 diattenuation. M3 diattenuation's persistent sampling and asymmetric posterior uncertainties, particularly in $H$, suggest residual coupling to other Mueller matrix parameters.

\begin{table*}[hbtp!]
\centering
\caption{Integrated autocorrelation times $\tau$ and saved chain length
divided by $\tau$ for the high-density-restart snapshots.  The saved lengths
before thinning were 352,892, 511,759, and 443,158 steps for $J$, $H$, and $K$, respectively.}
\label{tab:mcmc_tau}
\begin{tabular}{lrrrrrr}
\toprule
 & \multicolumn{2}{c}{$J$} & \multicolumn{2}{c}{$H$} & \multicolumn{2}{c}{$K$} \\
\cmidrule(lr){2-3}\cmidrule(lr){4-5}\cmidrule(lr){6-7}
Parameter & $\tau$ & $N/\tau$ & $\tau$ & $N/\tau$ & $\tau$ & $N/\tau$ \\
\midrule
Optics retardance & 3214.2 & 109.8 & 2887.9 & 177.2 & 1790.9 & 247.4 \\
Optics angle & 1561.4 & 226.0 & 3361.3 & 152.2 & 2208.2 & 200.7 \\
IMR retardance & 441.9 & 798.5 & 2119.6 & 241.4 & 537.4 & 824.7 \\
HWP retardance & 3248.4 & 108.6 & 6805.6 & 75.2 & 2199.3 & 201.5 \\
M3 diattenuation & 3844.0 & 91.8 & 9032.6 & 56.7 & 2944.6 & 150.5 \\
$\ln f$ & 193.3 & 1825.5 & 237.1 & 2158.4 & 180.5 & 2454.5 \\
\bottomrule
\end{tabular}
\end{table*}


All final corner plots are in Appendix \ref{app:corner_plots}. Pearson correlation coefficients, $r$, are used to quantify the linear relationships visible in the joint posteriors, with the sign of $r$ indicating the direction of the relationship and $|r|$ indicating its degree of linear association. In $J$, the optics retardance, HWP retardance, and M3 diattenuation exhibit pronounced positive tails, with the strongest correlations occurring between optics angle and HWP retardance ($r=-0.80$), optics retardance and HWP retardance ($r=0.68$), and optics retardance and M3 diattenuation ($r=0.62$). The $H$ posterior is particularly non-Gaussian as HWP retardance and M3 diattenuation are strongly anticorrelated ($r=-0.87$), while optics retardance and HWP retardance are positively correlated ($r=0.74$). The extended upper tail of M3 diattenuation produces both its asymmetric positive uncertainty and long autocorrelation time. In $K$, a narrow joint ridge strongly anticorrelates optics and HWP retardance ($r=-0.92$), accompanied by correlations between optics retardance and optics angle ($r=0.79$) and between optics angle and HWP retardance ($r=-0.72$). The $K$ optics angle posterior also has a long negative tail despite its compact central credible interval. 

Across all three bands, the marginalized IMR-retardance posterior is the closest to Gaussian and is less skewed than the other fitted parameters. The optics, HWP, and M3 posteriors are generally non-Gaussian, with asymmetric tails and substantial interparameter correlations. Although the directions and strengths of these correlations vary with wavelength, the dominant degeneracies are typically between the optics retardance and angle to the HWP retardance and M3 diattenuation.

\subsubsection{MCMC Residuals}
\label{subsubsec:mcmc_residuals}



\begin{table*}[hbtp!]
\centering
\caption{The root mean squared error (RMSE) and maximum absolute residuals for the MAP sample for JHK. The ranges in RMSE for HWP modulation sequences at each IMR angle are also reported.}
\label{tab:mcmc_residuals}
\begin{tabular}{lccc}
\toprule
Metric & $J$ & $H$ & $K$ \\
\midrule
RMSE & $1.946\times10^{-3}$ & $1.233\times10^{-3}$ & $5.036\times10^{-4}$ \\
Maximum absolute residual & $4.579\times10^{-3}$ & $3.640\times10^{-3}$ & $1.291\times10^{-3}$ \\
Per-IMR RMSE range & $(0.571$--$3.086)\times10^{-3}$ & $(0.433$--$1.808)\times10^{-3}$ & $(0.182$--$0.807)\times10^{-3}$ \\
\bottomrule
\end{tabular}
\end{table*}

Table~\ref{tab:mcmc_residuals} summarizes residuals at the joint MAP sample. The residuals are not purely random, as shown in Figure \ref{fig:JHK_random_chains}.  After averaging over every IMR angle, all three bands show a coherent pattern relative to HWP angle. The mean residual is positive near HWP angles $0^\circ$--$20^\circ$, negative near $30^\circ$--$60^\circ$, and closer to zero at the largest HWP angles.  Residuals decrease with wavelength, with the largest magnitude in $J$ and the smallest in $K$.  Residual scatter also depends on IMR angle. In J-band, the RMSE in $J$ ranges from $5.71\times10^{-4}$ at IMR = $90^\circ$ to $3.09\times10^{-3}$ at IMR = $0^\circ$.  

The correlation of the residual pattern with HWP angle points to possible missing components of the Mueller-matrix model. Several effects could contribute. The downstream diattenuations were fixed to zero in the present model. Although subtracting the mean single difference at each IMR angle removes any diattenuation contribution that is constant across the corresponding HWP sequence, it does not remove differential transmission that changes the amplitude or shape of the HWP modulation. Such behavior could consequently be absorbed partially into the fitted retardances or other model parameters. Beam walk from the HWP's rotation could also cause the exiting beam to sample regions of the IMR with different polarimetric properties. In addition, unequal background subtraction errors in the two Wollaston beams can bias the normalized single differences. If these errors vary during a HWP modulation sequence for a given IMR angle, they can persist even after the subtraction of the normalized single difference's mean offset. Uncertainty in the HWP or IMR angle zero points can similarly shift the phase of the predicted modulation, producing systematic residual structure near the modulation pattern's extrema. 

The efficacy of several possible solutions can be tested in future Mueller matrix models. Fitting the Mueller matrix model to double differences between complementary HWP angles can suppress beam-throughput and background offsets that remain stable between the paired measurements. Allowing the HWP, IMR, and optics diattenuations to vary as free parameters could account for HWP and IMR modulation dependent differential transmission. These changes will not correct for time variable background errors between complementary HWP angles (though the difference is expected to be small), beam walk across spatially nonuniform optics, or inaccurate angle zero points. Introducing additional parameters may also create new degeneracies and will increase the possibility of overfitting. Alternative background subtraction methods and additional measurements of the HWP and IMR angle zero points -- potentially with unpolarized standards with known DoLP rather than the dome light -- will therefore be needed to identify the root cause of the remaining residual structure.

\begin{figure*}[hbtp!]
    \centering
    \includegraphics[width=0.80\linewidth]
    {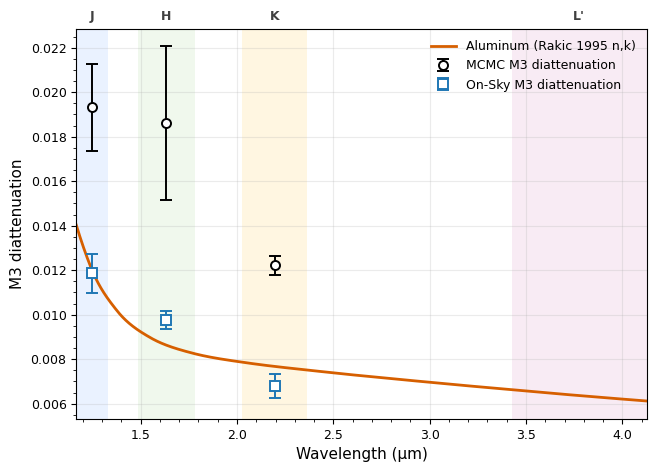}

    \caption{Comparison of the M3 diattenuation recovered from unpolarized standard stars with the aluminum Fresnel prediction and the dome-flat MCMC MAP values. The unpolarized standard measurements lie substantially closer to the Fresnel prediction, whereas the systematically larger dome flat estimates suggest that polarization in the incident dome illumination or other unmodeled calibration effects are absorbed into the fitted M3 parameter. This
    discrepancy motivates independently constraining the dome flat input polarization rather than assuming it is unpolarized. Shaded regions indicate the wavelength coverage of the three filters.}
    \label{fig:m3_diattenuation_comparison}
\end{figure*}

The fitted IMR solutions are close to half-wave in all three bands.  This is unexpected as the IMR was not optimized to act as a HWP in all these wavelength bands and was expected to have retardance significantly different from half a wave in at least some wavelength bins. This may be due to the IMR retardance being correlated with M3 diattenuation, the latter of which is also surprisingly far from Fresnel coefficient predictions (as per Fig. \ref{fig:m3_diattenuation_comparison}). More analysis of different forms of data like observations of standard stars with known polarizations are necessary to confirm whether the constant half-wave retardance of the IMR is a real physical property or a consequence of the modeling restrictions described above.

The retardance of the optics is small (less than 0.2 waves in all wavelength bands) with differing effective rotation angles in all three wavelengths. As there is currently no theoretical model of all the polarimetric effects of components between the IMR and Wollaston prism, it is unclear whether these are close to physical expectations or are absorbing other model inaccuracies and incompleteness. 

\begin{figure*}[hbtp!]
    \centering
    \includegraphics[width=\linewidth]{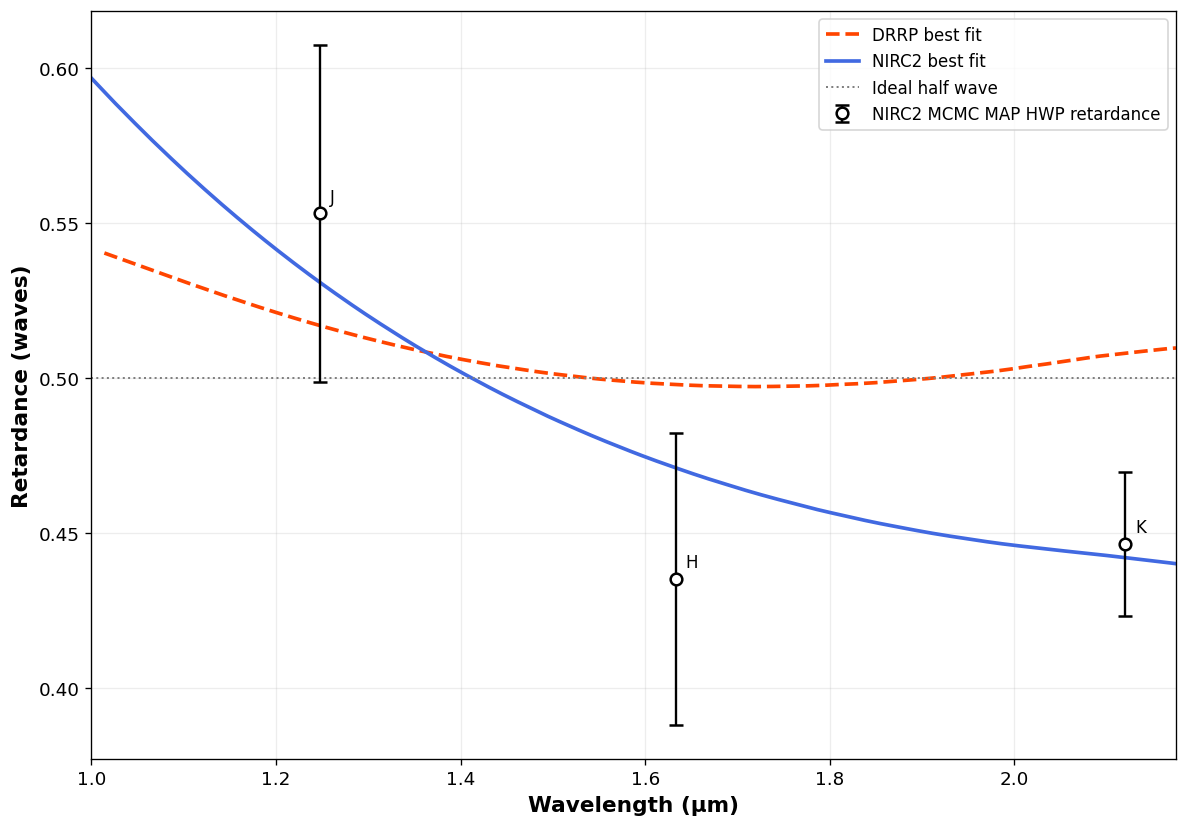}
    \caption{Comparison of laboratory measurements and MCMC MAP parameters for the HWP retardance in JHK. The red dashed line shows the laboratory best fit analytical model for nine wavelengths taken between 1--2~ $\mu$m, the blue solid line shows the analytical best fit model for the three broadband filter data points from MCMC MAP parameters values, and the black dashed line shows a perfect half-wave retardance. As the laboratory measurements isolate the HWP and sample its wavelength dependence more densely, we consider them the more reliable characterization. Future Mueller matrix fits will therefore fix the HWP retardance to the laboratory measurements.}
    \label{fig:HWP_Comparison}
\end{figure*}

There is some deviation between MCMC estimates and laboratory measurements of the HWP, as is shown in Fig.~\ref{fig:HWP_Comparison}. Fitting an analytical quartz and magnesium fluoride waveplate model directly to the three NIRC2 MCMC estimates produces nominal thicknesses of \(d_{\mathrm{MgF_2}}=800\pm200~\mu\mathrm{m}\) and \(d_{\mathrm{SiO_2}}=1000\pm300~\mu\mathrm{m}\), compared with \(1240\pm40~\mu\mathrm{m}\) and \(1570\pm50~\mu\mathrm{m}\) respectively from previous laboratory characterization \citep{melby2024half}. 
The manufacturer specifies thicknesses of $d_{\mathrm{MgF_2}}=1160\pm10~\mu\mathrm{m}$ and $d_{\mathrm{SiO_2}}=1480\pm10~\mu\mathrm{m}$ and a retardance of $\lambda/2\pm\lambda/70$ with an additional production tolerance of $\pm\lambda/300$. The DRRP thickness estimates agree with the specified thicknesses to within approximately $2\sigma$ and its $JHK$ retardances satisfy the combined retardance tolerance, whereas the MCMC MAP thicknesses are approximately $30\%$ smaller and the central MCMC MAP retardances fall outside the specified range. With only three broadband measurements for two fitted thicknesses, it is difficult to use the NIRC2 dome flat data to fit the physical thickness of either plate. In addition, these retardances were inferred as part of a full instrumental Mueller matrix fit and may absorb degeneracies with the IMR, static optics, unmodeled diattenuation, or other calibration systematics that have been mentioned in this section. 
In future analyses, we plan to fix the HWP retardance to the laboratory dual-rotating-retarder measurements represented by the red dashed curve in Fig. \ref{fig:HWP_Comparison}. Those measurements isolate the HWP without intervening NIRC2 optical components and densely sample its wavelength dependence, making them a more direct and robust characterization of the HWP itself.

The inferred M3 diattenuation decreases toward longer wavelengths qualitatively following the trend expected from the aluminum Fresnel
coefficients as shown in Fig. \ref{fig:m3_diattenuation_comparison}. However, the MCMC MAP values are all significantly larger than the nominal bare-aluminum predictions. The MAP diattenuations are $1.963\%$, $1.859\%$, and $1.222\%$ in $J$, $H$, and $K$, respectively, compared with theoretical values of $1.203\%$, $0.870\%$, and $0.771\%$.  The fitted values are approximately $1.6$, $2.1$, and $1.6$ times the corresponding Fresnel predictions. The discrepancy is larger than the plotted posterior uncertainties, with $H$ showing the greatest fractional difference, while the much narrower $K$-band posterior provides the most precise MCMC constraint. This trend and its implications are further discussed in Section \ref{subsec:m3_physical_model}.


\section{Unpolarized Standard Stars}
\label{subsec:unpolarized_selection}

Unpolarized standard stars are used to isolate instrumental polarization introduced by M3. For an intrinsically unpolarized source, any measured linear polarization after correction for downstream polarimetric effects can be attributed to M3. In NIRC2-Pol, these observations are also useful for comparing M3 diattenuation retrieved from dome flat calibrations and for identifying inaccuracies in the downstream Mueller matrix model. Deviations from the expected altitude-dependent sinusoidal pattern, including unexplained phase offsets, indicate possible errors in downstream retardance, angle offsets, or coordinate conventions.

The unpolarized standard-star sample is selected using several criteria. First, targets should have low intrinsic polarization in the near-infrared, ideally with literature measurements consistent with zero polarization at the $\lesssim0.1\%$ level. 
Secondly, targets with previous literature measurements with smaller uncertainties (preferably $\leq 0.1 \%$) are preferred as there is a lower likelihood of the standard's observed DoLP deviating significantly from zero. Finally, targets should preferably be observable over a broad range of telescope altitude during a single night.

The observations analyzed here were obtained during an observing night on 2026 July 6 UT and are summarized in
Table~\ref{tab:unpolarized_standards}. Only complete four-angle HWP cycles containing measurements at $0^\circ$, $22.5^\circ$, $45^\circ$, and $67.5^\circ$ were retained. The $K_p$ (central wavelength = 2.124 $\mu$m, bandpass width = 0.351 $\mu m$) observations were analyzed using the $K$-band (central wavelength = 2.196 $\mu$m, bandpass width = 0.336 $\mu m$) Mueller matrix calibration parameters. Future dome flat calibrations and science data will be measured in Kp. At present, using the $K$-band Mueller matrix for $K_p$ is unlikely to cause an issue due to the very similar central wavelengths and bandpasses.

\begin{table*}[hbtp!]
\centering
\caption{Unpolarized standard star observations used to determine the M3
diattenuation. The listed number of HWP cycles includes only complete
four-angle cycles retained by the analysis.}
\label{tab:unpolarized_standards}
\begin{tabular}{llccccc}
\toprule
Target & Band & Altitude Range & $t_{int}$ [s] & Coadds & HWP Cycles \\
\midrule
HD~168726 & $J$   &  $50.47^\circ$--$51.91^\circ$ & 0.50 & 20 & 8 \\
HD~164992 & $H$   & $39.13^\circ$--$42.28^\circ$ & 0.80 & 20 & 10 \\
HD~164992 & $K_p$ & $36.93^\circ$--$38.70^\circ$ & 0.80 & 20 & 5 \\
HD~239068 & $J$   & $42.89^\circ$--$47.96^\circ$ & 0.17 & 20--30 & 18 \\
HD~239068 & $H$   & $43.92^\circ$--$46.41^\circ$ & 0.17 & 20 & 12 \\
HD~239068 & $K_p$ & $45.00^\circ$--$47.48^\circ$ & 0.17 & 20 & 12 \\
\bottomrule
\end{tabular}
\end{table*}

The stellar photometry was measured with fixed circular apertures selected from background subtracted curves of growth. For each target, band, detector
shape, and Wollaston beam, the curve of growth was evaluated over aperture radii of 3--55 pixels. The aperture size selected was the smallest radius enclosing at least $95\%$ of the maximum measured curve-of-growth flux. The
median of the top and bottom beam radii was then adopted as a single fixed integer aperture radius for the corresponding target and band and applied to all individual exposures. The local background was estimated from an annulus with inner and outer radii of 65 and 85 pixels, respectively. 

\subsection{Determining the M3 Instrumental Polarization}
\label{subsec:m3_fit_method}

For each exposure, the normalized single difference between the two Wollaston beams was calculated using
Eq.~\ref{eq:normalized_difference}. The normalized double differences from each complete HWP cycle were calculated using Eq.~\ref{eq:q_normalized_double_difference} and Eq.~\ref{eq:u_normalized_double_difference}. The uncertainties on these double differences were propagated from the photometric uncertainties in the individual top and bottom beams.


A measurement matrix was constructed from the HWP, IMR, optics, and Wollaston Mueller matrices. This measurement matrix was used to retrieve the Stokes vector entering the HWP and exiting M3. For each configuration of the instrument, the Mueller matrix in Eq. \ref{eq:full_mueller_chain} but excluding M3 and on-sky coordinate rotations was evaluated using the HWP and IMR angles recorded for the corresponding exposure. The flux in the output beams for a given set of IMR and HWP angles $\theta$ is given by
\begin{equation}
    f_{b,\theta}
    =
    \boldsymbol{m}_{b,\theta}
    \boldsymbol{S}_{\rm post\text{-}M3},
    \label{eq:modeled_beam_flux}
\end{equation}
where $b$ corresponds to the top or bottom beam,
$\boldsymbol{m}_{b,\theta}$ is the first row of
the Mueller matrix. The four elements in $\boldsymbol{m}_{b,\theta}$ describe how incident $I$, $Q$, $U$, and $V$ contribute to the intensity recorded by that Wollaston beam.

For the normalized post-M3 Stokes vector $\boldsymbol{s}_{\rm post\text{-}M3}$, the Mueller matrix for the normalized single difference at HWP and IMR angle combination $\theta$ is expressed
as
\begin{equation}
    ND_{\rm model}^{\theta}
    =
    \boldsymbol{r}_{\theta}
    \boldsymbol{s}_{\rm post\text{-}M3},
    \qquad
    \boldsymbol{r}_{\theta}
    =
    \frac{
        \boldsymbol{m}_{{\rm top},\theta}
        -
        \boldsymbol{m}_{{\rm bottom},\theta}
    }{
        m_{{\rm top},\theta,I}
        +
        m_{{\rm bottom},\theta,I}
    },
    \label{eq:normalized_model_row}
\end{equation}
where $m_{b,\theta,I}$ is the first element of
$\boldsymbol{m}_{b,\theta}$ and gives the $I\rightarrow I$ response (throughput) of that beam. 

Using $r_{\theta_{HWP},j}$ ($j\in\{I,Q,U,V\}$) to denote the Mueller matrix coefficient for a specific critical HWP angle, the two normalized double differences form the measurement
system
\begin{equation}
    \underbrace{
    \begin{pmatrix}
        y_Q\\
        y_U
    \end{pmatrix}}_{\boldsymbol{y}}
    =
    \underbrace{
    \frac{1}{2}
    \begin{pmatrix}
        r_{0^\circ,I}-r_{45^\circ,I}
        &
        r_{0^\circ,Q}-r_{45^\circ,Q}
        &
        r_{0^\circ,U}-r_{45^\circ,U}
        &
        r_{0^\circ,V}-r_{45^\circ,V}
        \\
        r_{22.5^\circ,I}-r_{67.5^\circ,I}
        &
        r_{22.5^\circ,Q}-r_{67.5^\circ,Q}
        &
        r_{22.5^\circ,U}-r_{67.5^\circ,U}
        &
        r_{22.5^\circ,V}-r_{67.5^\circ,V}
    \end{pmatrix}}_{\boldsymbol{A}}
    \begin{pmatrix}
        1\\
        q\\
        u\\
        v
    \end{pmatrix}
    \label{eq:double_difference_measurement_matrix}
\end{equation}
Here, $y_Q$ is the normalized double difference formed from the $0^\circ$ and $45^\circ$ HWP measurements, while $y_U$ is formed from the $22.5^\circ$ and $67.5^\circ$ measurements. 

Circular polarization was assumed to be negligible, meaning $v=0$. Separating the
intensity column of $\boldsymbol{A}$ from its $q$ and $u$ columns gives
\begin{equation}
    \boldsymbol{y}
    =
    \underbrace{
    \begin{pmatrix}
        a_{Q,I}\\
        a_{U,I}
    \end{pmatrix}}_{\boldsymbol{a}_{I}}
    +
    \underbrace{
    \begin{pmatrix}
        a_{Q,Q} & a_{Q,U}\\
        a_{U,Q} & a_{U,U}
    \end{pmatrix}}_{\boldsymbol{X}}
    \begin{pmatrix}
        q\\
        u
    \end{pmatrix}.
    \label{eq:reduced_double_difference_system}
\end{equation}
The vector $\boldsymbol{a}_{I}$ describes leakage from total intensity into the two double differences, while $\boldsymbol{X}$ describes the response of double difference measurements to incident $q$ and $u$. The recovered normalized linear Stokes parameters are therefore
\begin{equation}
    \begin{pmatrix}
        \widehat{q}\\
        \widehat{u}
    \end{pmatrix}
    =
    \boldsymbol{X}^{+}
    \left(
        \boldsymbol{y}
        -
        \boldsymbol{a}_{I}
    \right),
    \label{eq:unpolarized_matrix_inversion}
\end{equation}
where the superscript $+$ denotes the Moore--Penrose pseudoinverse.

Holding the model measurement matrix fixed and treating the uncertainties in
$y_Q$ and $y_U$ as independent, the propagated covariance of the recovered
Stokes parameters was
\begin{equation}
    \boldsymbol{C}_{qu}
    =
    \boldsymbol{X}^{+}
    \operatorname{diag}
    \left(
        \sigma_{y_Q}^{2},
        \sigma_{y_U}^{2}
    \right)
    \left(\boldsymbol{X}^{+}\right)^{\mathsf T}.
    \label{eq:unpolarized_stokes_covariance}
\end{equation}
This covariance propagates the photometric uncertainties in the measured
double differences but does not include uncertainty in the fixed Mueller matrix parameters. No additional constant $q$ or $u$ offset was fitted after the HWP pair subtraction. Before reflection from M3, the stellar input was assumed
to be unpolarized.

The matrix-inverted $q$ and $u$ measurements were fitted with the physical M3 model, with the altitude and M3 Mueller matrices taking in the rotation or diattenuation parameters respectively as free inputs:
\begin{equation}
    \boldsymbol{S}_{\rm model}(a)
    =
    \boldsymbol{M}_{\rm alt}
    \left(a+\Delta a\right)
    \boldsymbol{M}_{\rm M3}
    (\epsilon_{\rm M3})
    \boldsymbol{S}_{\rm in}
    \label{eq:m3_altitude_model}
\end{equation}
where $a$ is the recorded telescope altitude, $\Delta a$ is an additive altitude angle offset, and  $\epsilon_{\rm M3}$ is the M3 diattenuation. 

A single $\epsilon_{\rm M3}$ and $\Delta a$ was fitted for all standards within each wavelength band. These parameters were determined by minimizing the joint weighted residuals in $q$ and $u$, with $-0.20\leq\epsilon_{\rm M3}\leq0.20$ and
$-45^\circ\leq\Delta a\leq45^\circ$. The reported uncertainties were estimated from the local numerical Jacobian and scaled by the reduced $\chi^2$. The resulting parameters and fit diagnostics are listed in
Table~\ref{tab:m3_fits}.

\begin{table*}[t]
\centering
\caption{M3 parameters fitted to the matrix inverted unpolarized standard
measurements. Diattenuation is reported as a dimensionless fraction.}
\label{tab:m3_fits}
\begin{tabular}{lccc}
\toprule
Quantity & $J$ & $H$ & $K_p$ \\
\midrule
M3 diattenuation $\epsilon_{\rm M3}$
    & $0.0119\pm0.0009$
    & $0.0098\pm0.0004$
    & $0.0068\pm0.0005$ \\
Altitude offset $\Delta a$ [$^\circ$]
    & $8 \pm 2 $
    & $6 \pm 1 $
    & $-4 \pm 3$ \\
Number of $q,u$ measurements
    & 52 & 44 & 34 \\
Residual RMSE
    & $0.00379$
    & $0.00234$
    & $0.00306$ \\
Reduced $\chi^2$
    & 2.06 & 4.04 & 2.91 \\
\bottomrule
\end{tabular}
\end{table*}

The fitted model reproduces the overall altitude dependent instrumental polarization and wavelength dependence of the recovered polarization, showing that the Mueller matrix inversion preserves the expected modulation in $q$ and $u$ due to varying altitude. However, the reduced $\chi^2$ values are greater than one in every band, indicating that
the remaining scatter is larger than expected from the propagated photometric uncertainties alone. 

The positive altitude offsets in $J$ and $H$ are inconsistent with zero at the level of their formal uncertainties, whereas the $K_p$ offset is
much closer to zero. An actual error in the telescope altitude zero point would be expected to be wavelength independent. The wavelength dependence of the fitted altitude offsets is therefore more plausibly interpreted as
effective phase corrections that absorb residual errors in the downstream retardances, HWP or IMR zero points, or small intrinsic polarization from the unpolarized standards, with the former two being much more likely. 

\begin{figure*}[hbtp!]
    \centering
    \includegraphics[width=\linewidth]
    {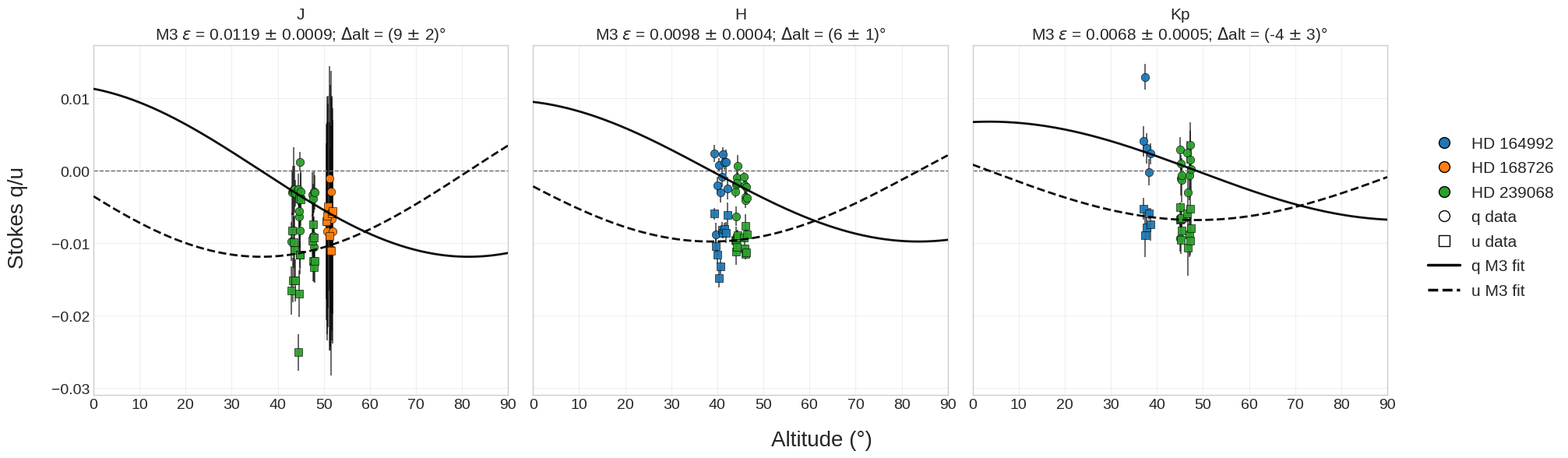}
    \caption{Matrix-inverted Stokes $q$ and $u$ for the unpolarized standards as a function of telescope altitude. Different colors identify the
    individual standards, while circles and squares denote $q$ and $u$, respectively. The solid ($q$) and dashed ($u$) curves show the corresponding best fitting M3 model in each wavelength band.}
    \label{fig:m3_unpolarized_fits}
\end{figure*}

\subsection{Comparison with a Physical Mirror Model}
\label{subsec:m3_physical_model}

As a consistency check, the on-sky M3 diattenuations are compared with both the dome flat MCMC results and a physical model based on the Fresnel reflection coefficients of aluminum (the known material of M3). For a metallic mirror, the complex refractive index and angle of incidence determine the reflected intensities of the $s$ and $p$ polarization components. The predicted diattenuation is
\begin{equation}
    \epsilon
    =
    \frac{|r_s|^2-|r_p|^2}
         {|r_s|^2+|r_p|^2},
    \label{eq:fresnel_diattenuation}
\end{equation}
where $\epsilon$ is the dimensionless diattenuation predicted for M3, and $r_s$ and $r_p$ are the complex Fresnel amplitude reflection coefficients for electric fields polarized perpendicular ($s$) and parallel ($p$) to the plane of incidence. $|r_s|^2$ and $|r_p|^2$ are the corresponding intensity reflectances. With this convention, positive $\epsilon_{\rm model}$ indicates that M3 reflects $s$-polarized light more efficiently than $p$-polarized light. The reflection coefficients depend on wavelength, angle of incidence, and the complex refractive index of the mirror coating. For an initially unpolarized source, the magnitude of diattenuation is the fractional linear polarization introduced by M3.


\begin{table*}[hbtp!]
\centering
\caption{Comparison of the theoretical aluminum Fresnel diattenuation, the M3 diattenuation inferred from the dome-flat MCMC calibration, and the independent on-sky unpolarized-standard fit. The $K_p$ on-sky measurements were inverted using the $K$-band calibration.}
\label{tab:m3_physical_comparison}
\begin{tabular}{lccc}
\toprule
Diattenuation estimate & $J$ & $H$ & $K$/$K_p$ \\
\midrule
Aluminum Fresnel model
    & 0.0120 & 0.0087 & 0.0077 \\
Dome-flat MCMC MAP
    & $0.0197_{-0.0005}^{+0.001}$ & $0.0186_{-0.0004}^{+0.002}$ & $0.0122_{-0.0001}^{+0.0003}$ \\
On-sky unpolarized standards
    & $0.0119\pm0.0009$
    & $0.0098\pm0.0004$
    & $0.0068\pm0.0005$ \\
\bottomrule
\end{tabular}
\end{table*}

The M3 diattenuation values derived from on-sky unpolarized standards follow the predicted decrease in M3 diattenuation with increasing wavelength and are substantially closer to the bare aluminum Fresnel predictions than the dome flat MCMC MAP values. The $J$-band measurement agrees particularly closely with the theoretical value, while the $H$-band result is slightly higher and the $K_p$ result slightly lower. In contrast, the dome flat MCMC values are systematically larger in all three bands. This difference suggests that the M3 parameter in the dome flat fit may absorb part of the downstream model mismatch, including neglected diattenuation or degeneracies among the fitted retardances. It may also indicate that the dome flat illumination is already polarized before reflection from M3, in which case the dome flat fit would incorrectly attribute the combined incident and M3-induced polarization to M3 alone. 

\begin{figure*}[hbtp!]
    \centering
    \includegraphics[width=\linewidth]
    {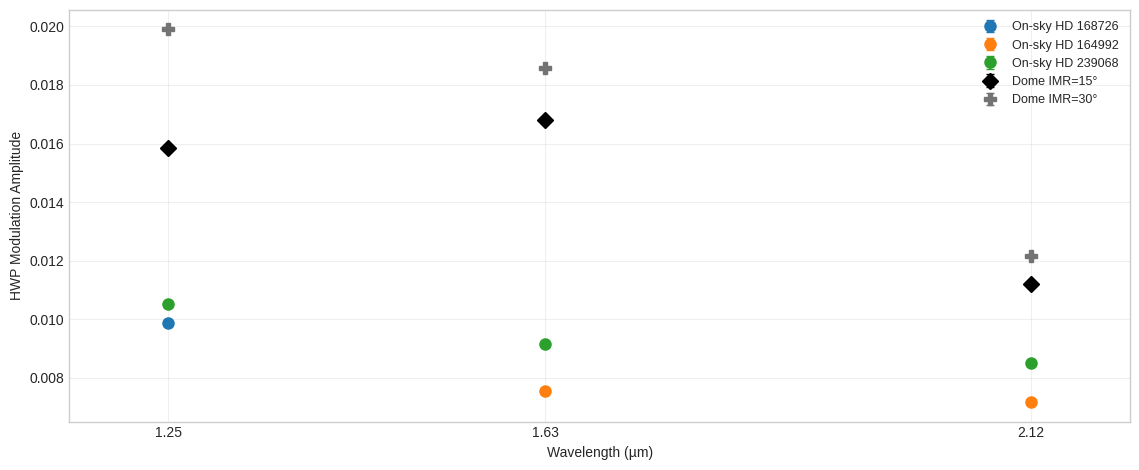}

    \caption{Comparison of the HWP modulation amplitude measured from unpolarized standard stars and dome flats in the $J$, $H$, and $K_p$ bands. Each sequence was fitted with $SD(\theta_{\rm HWP})=c_0+A\sin[4(\theta_{\rm HWP}-\theta_0)]$, and the fitted amplitude $A$ is shown. The dome flat sequences at fixed IMR angles of $15^\circ$ and $30^\circ$ (the closest IMR angles compared to the on-sky data) exhibit systematically larger modulation than the on-sky standards, suggesting that the incident dome illumination may be polarized before M3 or there are additional unmodeled calibration systematics.}
    \label{fig:HWP_modulation_amplitude}
\end{figure*}

Fig.~\ref{fig:HWP_modulation_amplitude} compares the amplitude of the HWP modulation measured from the dome flats and unpolarized standards. This comparison provides a model independent test of whether the elevated M3 diattenuation inferred from the dome flats corresponds to a larger input polarization in the data, as would be expected if the dome illumination were already polarized before reaching M3. For each combination of on-sky target and wavelength band, each selected HWP modulation sequence with dome flats and fixed IMR angle, the normalized single differences were fitted directly using weighted nonlinear least squares with
\begin{equation}
    SD(\theta_{\rm HWP})
    =
    c_0+
    A\sin\left[4\left(\theta_{\rm HWP}-\theta_0\right)\right],
    \label{eq:single_sine_hwp_modulation}
\end{equation}
where $c_0$ is the mean single-difference offset, $A\geq0$ is the fitted sinusoidal amplitude, and $\theta_0$ is the fitted HWP phase offset. As the modulation is periodic in $\theta_0$ with a period of $90^\circ$, the phase was restricted to $-45^\circ\leq\theta_0<45^\circ$. Several initial phase values were tested for each series, and the solution with the lowest $\chi^2$ was retained. The plotted quantity in Figure \ref{fig:HWP_modulation_amplitude} is the amplitude, $A$, with its uncertainty is $\sigma_A$ and is obtained from the covariance of the three parameter fit of $c_0$, $A$ and $\theta_0$. These uncertainties use photometric errors.

A constant error in the adopted HWP angle zero point is strongly degenerate with $\Delta a$ as both rotate the recovered M3-induced polarization
pattern in the $q$--$u$ plane. For an ideal HWP, a HWP fast-axis error $\delta\theta_{\rm HWP}$ produces the same phase displacement as an altitude
offset with $\Delta a = 2\delta\theta_{\rm HWP}$. The nonzero fitted values of $\Delta a$ therefore indicate a phase mismatch and could be explained with
incorrect HWP offsets, but they cannot distinguish that possibility from IMR angle or optics rotation angle errors.

The fitted dome flat amplitudes range from approximately $0.0112$ to $0.0199$, whereas the on-sky amplitudes range from approximately $0.0072$ to $0.0106$. The dome flat modulation is therefore larger than the on-sky modulation at every wavelength shown. This is consistent with an additional polarized component in the dome illumination before M3, which could cause the dome-flat MCMC analysis to infer an artificially large effective M3 diattenuation when the incident dome light is assumed to be unpolarized. 

Nevertheless, the elevated reduced $\chi^2$ values and effective altitude offsets in the on-sky fits show that the agreement with the Fresnel curve should not be interpreted as a complete validation of the current Mueller matrix. Improved altitude coverage is needed to quantify the difference between the theoretical Fresnel and measured M3 diattenuation.

\section{Conclusions \& Future Work}
\label{sec:conclusions}

We have presented preliminary $JHK$ Mueller matrix characterization of the new Keck/NIRC2 Polarimetry mode (NIRC2-Pol). The present model describes the tertiary/Nasmyth mirror (M3), the half-wave plate (HWP), the image rotator (IMR), static optics between the IMR and Wollaston prism, and the Wollaston analyzer. Model parameters were constrained using dome flat sequences spanning ten HWP angles and nine IMR angles in each band and on-sky unpolarized standards spanning $\sim 10^{\circ}$ altitude. 

The dome flat Markov Chain Monte Carlo (MCMC) analysis revealed important parameter degeneracies. The maximum-a-posteriori (MAP) IMR retardances are $0.516_{-0.008}^{+0.008}$, $0.476_{-0.007}^{+0.009}$, and $0.483_{-0.005}^{+0.005}$ waves in $J$, $H$, and $K$, respectively, placing all three solutions unexpectedly close to half-wave retardance. The fitted HWP retardances are $0.54_{-0.02}^{+0.04}$, $0.44_{-0.05}^{+0.02}$, and $0.45_{-0.02}^{+0.02}$ waves and differ from the more densely sampled laboratory dual-rotating-retarder polarimeter measurements. The optics, HWP, and M3 posteriors are generally non-Gaussian and correlated, whereas the IMR retardance posteriors are closer to Gaussian. The dome flat M3 MAP diattenuations are $0.0197_{-0.0005}^{+0.001}$, $0.0186_{-0.0004}^{+0.002}$, and $0.0122_{-0.0001}^{+0.0003}$ in $J$, $H$, and $K$, substantially larger than the corresponding bare aluminum Fresnel predictions of $0.0120$, $0.0087$, and $0.0077$. These dome flat estimates should therefore be interpreted as effective parameters within the restricted model rather than robust physical measurements of M3.

The MAP models reproduce the dominant HWP and IMR-dependent modulation, with normalized single difference RMSE values of $1.95\times10^{-3}$, $1.23\times10^{-3}$, and $5.04\times10^{-4}$ in $J$, $H$, and $K$, respectively. However, the residuals contain coherent structure with HWP angle and vary with IMR angle. This structure indicates that the fitted additive scatter does not represent purely random noise and that the current Mueller matrix is likely incomplete. Possible contributors to systematic residuals include downstream diattenuation being fixed to zero, HWP beam walk, imperfect background subtraction, offset angle errors, and the limitations of 
subtracting the mean single difference offsets within each IMR angle. The reported dome flat posterior intervals are consequently conditional on these modeling assumptions and on neglecting the covariance introduced by subtracting a common mean from each IMR sequence.

Unpolarized standard star observations provide a more direct estimate of M3's properties. After inverting the measurement matrix representing the downstream $JHK$ Mueller matrices, the fitted M3 diattenuations are $0.0119\pm0.0009$, $0.0098\pm0.0004$, and $0.0068\pm0.0005$ in $J$, $H$, and $K_p$ respectively. These values follow the expected decrease with wavelength and are much closer to the Fresnel predictions for aluminum than the dome flat MCMC values. The corresponding altitude offsets are $8\pm2^\circ$, $6\pm1^\circ$, and $-4\pm3^\circ$, while the reduced $\chi^2$ values are 2.06, 4.04, and 2.91. The wavelength dependence of the offsets and the excess residual scatter indicate that they are more likely effective corrections for remaining downstream model Mueller matrix inaccuracies than physical errors in the telescope altitude zero point.

The systematically larger M3 diattenuations and HWP modulation amplitudes measured from the dome flats (compared to measurements from unpolarized standards) are consistent with either additional polarization in the dome illumination before M3 or other unmodeled differences between the two measurements. If the dome illumination is polarized, assuming an unpolarized input would cause that signal to be absorbed into the fitted M3 and downstream parameters. This discrepancy demonstrates that the assumption that the input polarization state is unpolarized must be removed and input polarization DoLP and AoLP fitted as free parameters or calculated analytically before the dome flat parameters can be re-adopted for calibrating on-sky measurements.

These results establish an initial framework for modeling the dominant instrumental polarization and cross-talk in NIRC2-Pol, but they do not yet establish the accuracy with which astrophysical Stokes parameters can be recovered. Polarimetric accuracy will be established in future work after adjustments to the present Mueller matrix model. Additionally, the present quantitative calibration is limited to $JHK$ and does not include $L'$. Validation with polarized standards and science targets also remains necessary. The current model should therefore be regarded as preliminary rather than the final calibration for performing quantitative polarimetry on science targets.


The immediate priority is to revise the $JHK$ Mueller-matrix fit using stronger independent constraints on the physical components. In particular, the HWP retardance can be fixed to the laboratory dual-rotating retarder polarimeter measurements, which isolate the HWP and sample its wavelength dependence more densely than the three broadband dome flat measurements. This will reduce the ability of the HWP term to absorb errors associated with the IMR, static optics, M3, or other unmodeled effects. The resulting fits could potentially test whether the inferred near half-wave IMR retardances persist after the HWP response is externally constrained.

Future fits will also test nonzero HWP, IMR, and static optics diattenuations, as well as a non-unity Wollaston throughput ratio. Rather than using single differences to perform fits of downstream Mueller matrix parameters, double differences between HWP angles can also be used. The latter approach is the observable used for on-sky matrix inversion and is also much more robust against diattenuation and flat-fielding effects.

From the significant difference in amplitude of single difference modulation between dome flat and on-sky data, future fits will also include fitted or analytically calculated input DoLP and AoLP before M3, with additional polarization potentially induced by reflection of the calibration lamps off of the dome. A linear polarizer placed upstream of the HWP would also provide the most direct way to inject known polarization states and break degeneracies among M3, the HWP, the IMR, optics, and Wollaston transmission.




Additional unpolarized standards will be observed over substantially broader ranges of telescope altitude. This will help distinguish a true modulation pattern induced by M3's diattenuation from downstream polarimetric effects. Polarized standards with reliable near-infrared measurements, in the absence of a linear polarizer, are required to determine the absolute accuracy of the recovered degree and angle of linear polarization. Such standards have already been observed across a range of AoLP and telescope altitudes; processing these data can serve as validation of the complete Mueller matrix model. All on-sky calibration procedures will also be extended to $L'$. 

Finally, the validated Mueller matrix will be applied to representative polarized science targets and compared with reductions that assume an ideal instrument. For circumstellar disks, useful quantitative diagnostics include any changes in polarized intensity, AoLP, polarimetric morphology, and residual signal in $U_\phi$ (azimuthal cross-polarization). Together, these steps will move NIRC2-Pol from a qualitative polarimetric mode with a preliminary model to an infrared polarimeter capable of quantitative polarimetry for a multitude of science cases.

\appendix

\section{MCMC Corner Plots}
\label{app:corner_plots}

This appendix presents the complete corner plots for the MCMC analyses in the $J$, $H$, and $K$ bands. These plots complement the parameter estimates reported in the main text by showing both the marginalized posterior distributions and the joint distributions between parameter pairs. The inclusion of these plots is to show non-Gaussian posterior shapes, asymmetric uncertainties, parameter correlations, and degeneracies that cannot be conveyed fully by the median, MAP value, and posterior uncertainty intervals alone.

\begin{figure*}[hbtp!]
    \centering
    \includegraphics[width=\textwidth]
    {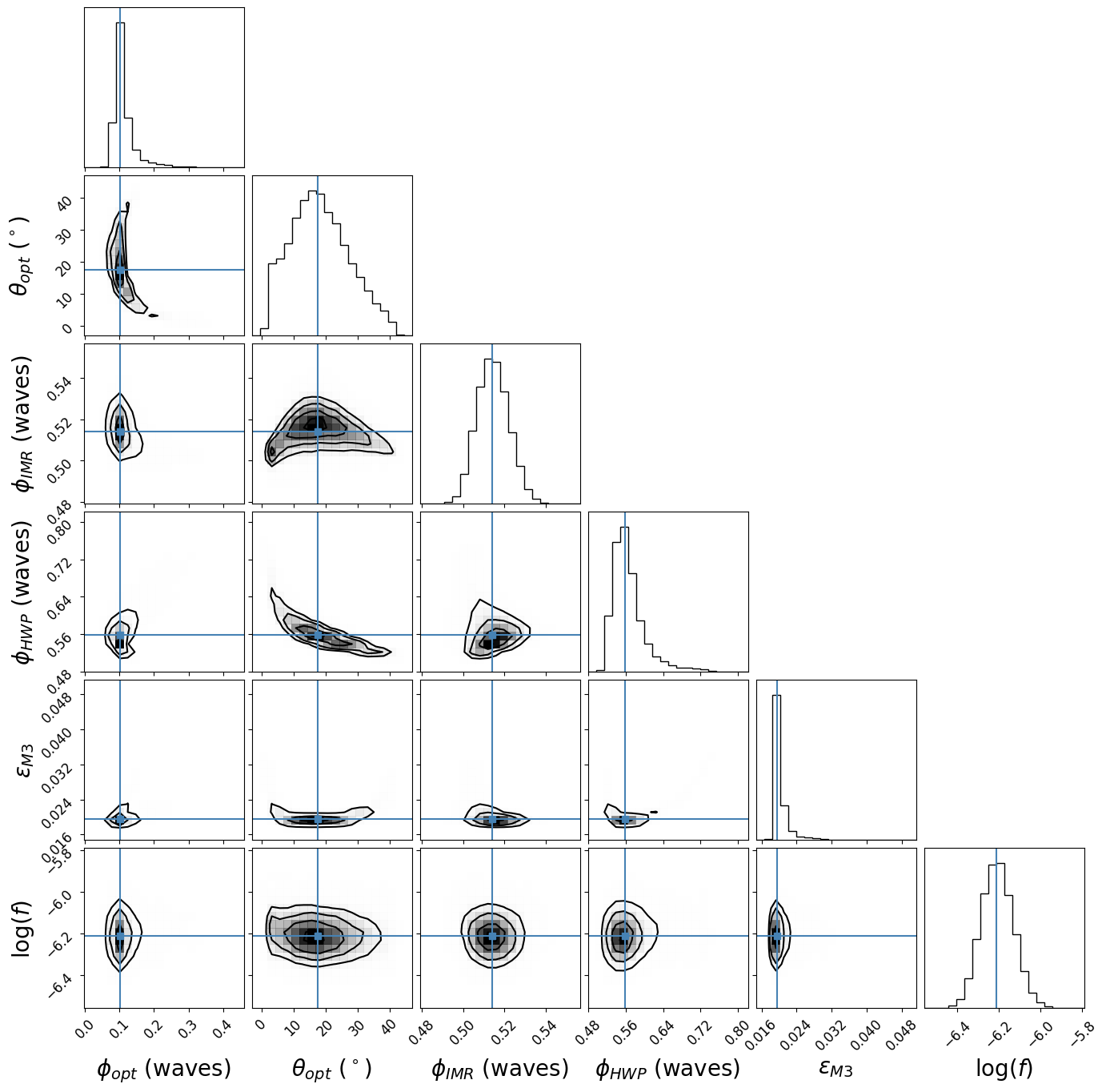}

    \caption{Corner plot of the post burn-in, thinned samples from the high-likelihood restarted MCMC analysis in $J$ band. Diagonal panels show
    marginalized posterior distributions, while off-diagonal panels show the corresponding joint posteriors. The most prominent features are positive tails in the optics retardance, HWP retardance, and M3 diattenuation, and correlation between the HWP retardance and both the optics angle and optics retardance.}
    \label{fig:j_corner_plot}
\end{figure*}

\begin{figure*}[hbtp!]
    \centering
    \includegraphics[width=\textwidth]
    {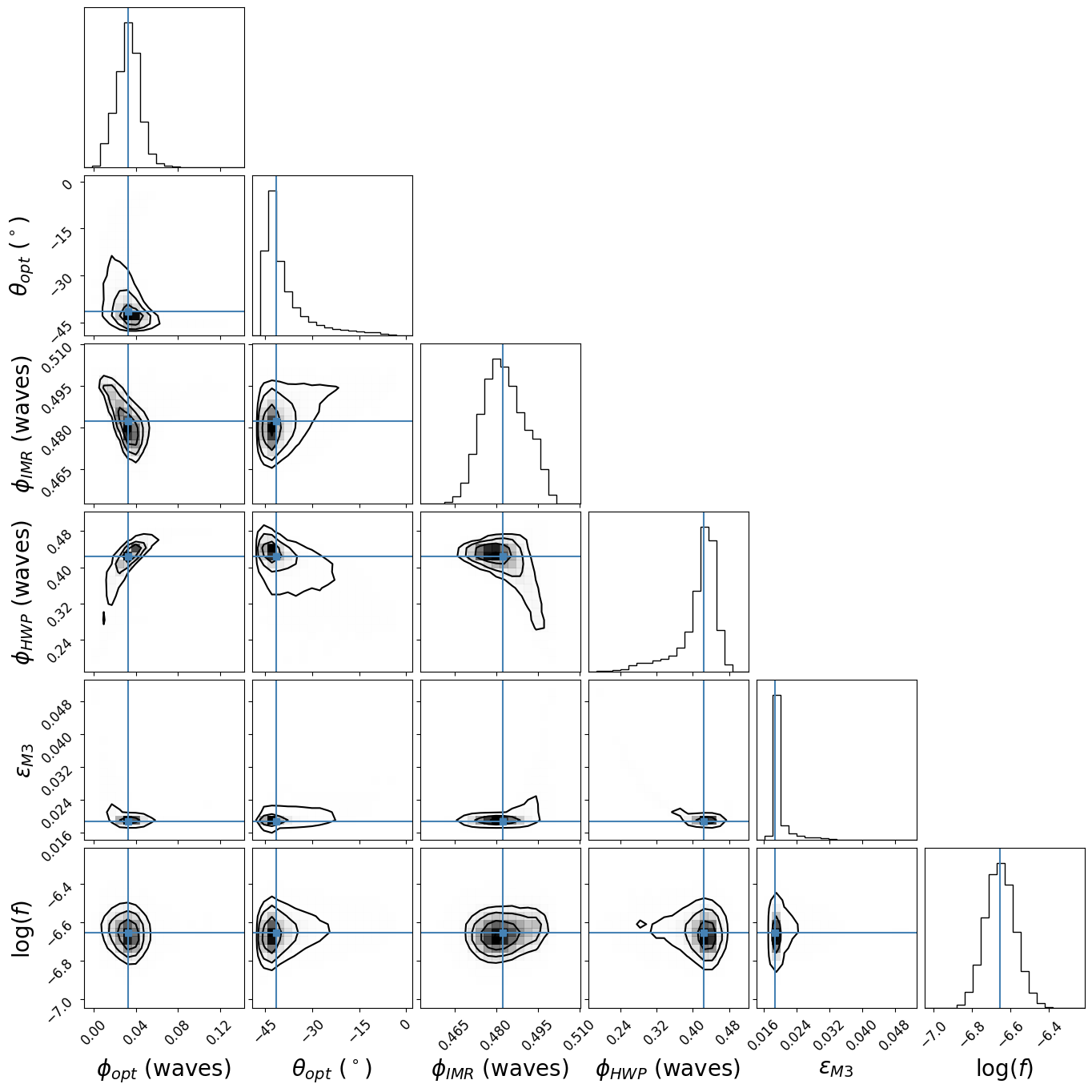}

    \caption{Corner plot of the post burn-in, thinned samples from the high-likelihood restart MCMC analysis in $H$ band. Diagonal panels show
    marginalized posterior distributions, while off-diagonal panels show the corresponding joint posteriors. Joint posteriors are notably non-Gaussian, with a strong anticorrelation between HWP retardance and M3 diattenuation and an extended upper tail in M3 diattenuation.}
    \label{fig:h_corner_plot}
\end{figure*}

\begin{figure*}[hbtp!]
    \centering
    \includegraphics[width=\textwidth]
    {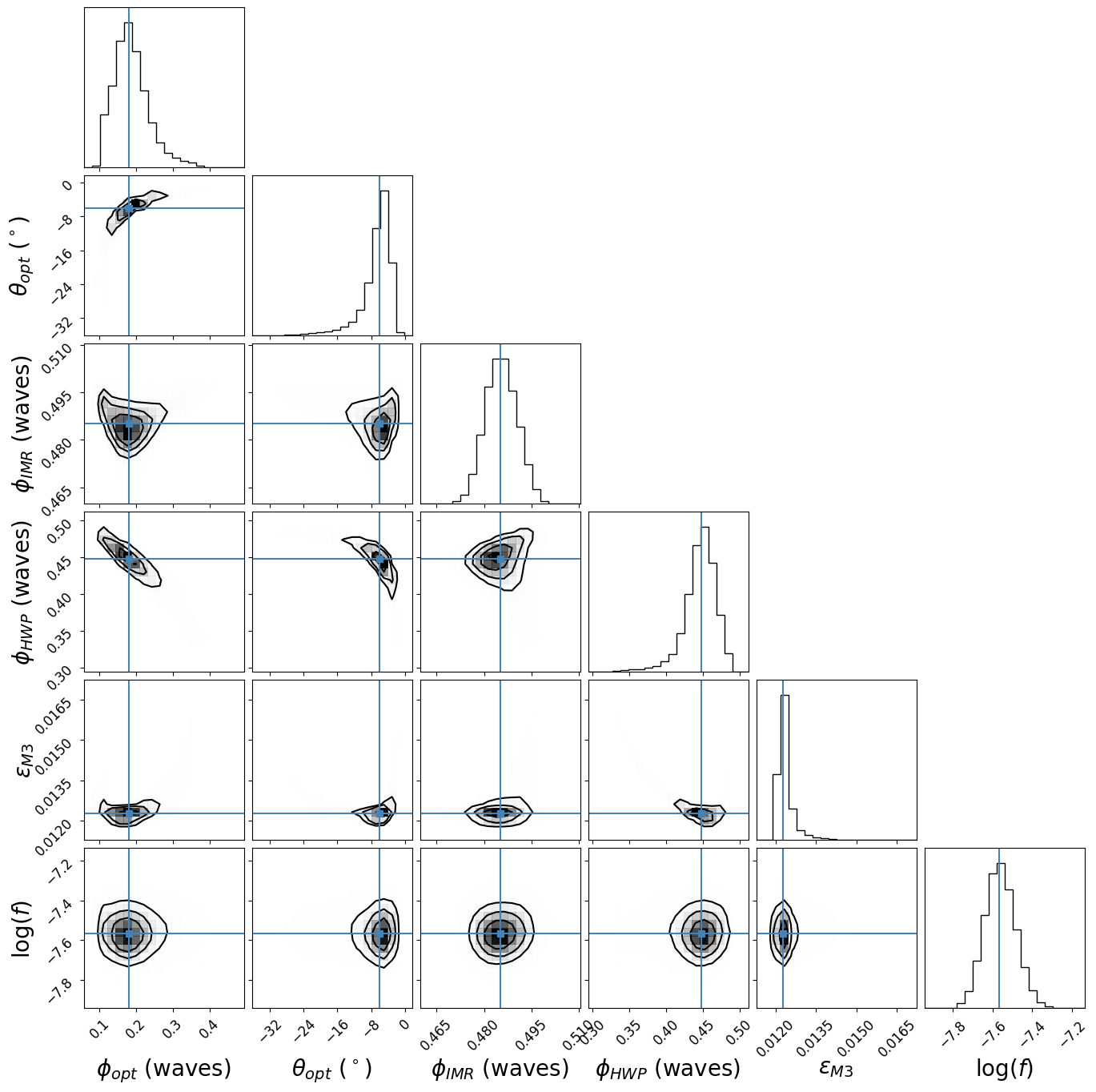}

    \caption{Corner plot of the post burn-in, thinned samples from the high-likelihood restart MCMC analysis in $K$ band. Diagonal panels show
    marginalized posterior distributions, while off-diagonal panels show the corresponding joint posteriors. A narrow ridge strongly couples the
    optics and HWP retardances, while the optics angle is correlated with both quantities and exhibits an extended negative tail.}
    \label{fig:k_corner_plot}
\end{figure*}





\section*{Acknowledgments}

Some of the data presented herein were obtained at Keck Observatory, which is a private 501(c)3 non-profit organization operated as a scientific partnership among the California Institute of Technology, the University of California, and the National Aeronautics and Space Administration. The Observatory was made possible by the generous financial support of the W. M. Keck Foundation. The authors wish to recognize and acknowledge the very significant cultural role and reverence that the summit of Maunakea has always had within the Native Hawaiian community. We are most fortunate to have the opportunity to conduct observations from this mountain.

This material is based upon work supported by the National Science Foundation Astronomy \& Astrophysics Postdoctoral Fellowship Award No. 2401654 for author BLL. Any opinions, findings, and conclusions or recommendations expressed in this material are those of the authors and do not necessarily reflect the views of the National Science Foundation. This work was also supported by the Mt. Cuba Astronomical Foundation and the University of California Observatories Mini-Grant Program. J.N.A was supported by NASA through the NASA Hubble Fellowship grant \#HST-HF2-51547.001-A awarded by the Space Telescope Science Institute, which is operated by the Association of Universities for Research in Astronomy. JL, CAC, and MF acknowledge support from the Heising-Simons Foundation under grant No. 2022-354 and from the National Science Foundation under grants No. 1909641. This research made use of Photutils, an Astropy package for detection and photometry of astronomical sources \cite{Bradley2025-oh}.


\newpage
\bibliography{report}
\bibliographystyle{spiebib}

\end{document}